\documentclass[journal]{IEEEtran}

\usepackage{amsmath,bm}
\usepackage{kantlipsum}
\usepackage{ragged2e}
\usepackage{relsize}
\usepackage[dvips]{graphicx}
\usepackage{algorithm}
\usepackage{algorithmicx}
\usepackage{algpseudocode}
\usepackage{amsfonts}
\usepackage{cite}
\usepackage{dsfont}
\usepackage{amssymb}
\usepackage{color}
\usepackage{tabls}
\usepackage{slashbox}
\usepackage{stfloats}
\usepackage[table,xcdraw]{xcolor}
\usepackage{hhline}
\usepackage{tabu}
\usepackage{cases}
\usepackage[caption=false]{subfig}
\usepackage[breaklinks]{hyperref}
\newtheorem{thm} {Theorem}
\newtheorem{asump} {Assumption}
\newtheorem{cond} {Condition}

\newtheorem{lemma} {Lemma}
\newtheorem{examp} {Example}

\newcommand*{\QEDB}{\hfill\ensuremath{\square}}
\newcommand{\tabincell}[2]{\begin{tabular}{@{}#1@{}}#2\end{tabular}}

\renewcommand\IEEEkeywordsname{Index Terms}
\let\oldIEEEkeywords\IEEEkeywords
\def\IEEEkeywords{\oldIEEEkeywords\normalfont\bfseries\ignorespaces}

\makeatletter
\newcommand{\thickhline}{
    \noalign {\ifnum 0=`}\fi \hrule height 1pt
    \futurelet \reserved@a \@xhline
}
\makeatother

\makeatletter
\newcommand{\tthickhline}{
	\noalign {\ifnum 0=`}\fi \hrule height 1.5pt
	\futurelet \reserved@a \@xhline
}
\makeatother

\makeatletter
\newcounter{parentalgorithm}

\makeatother

\begin{document}

\title{Distributed Learning for Stochastic Generalized Nash Equilibrium Problems}

\author{Chung-Kai~Yu,~\IEEEmembership{Student Member,~IEEE,}
        Mihaela~van~der~Schaar,~\IEEEmembership{Fellow,~IEEE,}
        and~Ali~H.~Sayed,~\IEEEmembership{Fellow,~IEEE}
\thanks{The authors are with the Department of Electrical Engineering, University of California, Los Angeles, CA 90095-1594 USA (e-mail: ckyuna@ucla.edu, \{mihaela,sayed\}@ee.ucla.edu). }
\thanks{This work was supported in part by NSF grants  ECCS-1407712 and CCF-1524250, and by an ONR Mathematical Data Sciences grant. An early short version of this work appeared in the conference publication~\cite{Yu16}.}
}

\markboth{Journal of \LaTeX\ Class Files,~Vol.~11, No.~4, December~2012}
{Shell \MakeLowercase{\textit{et al.}}: Bare Demo of IEEEtran.cls for Journals}

\maketitle

\begin{abstract}
This work examines a stochastic formulation of the generalized Nash equilibrium problem (GNEP) where agents are subject to randomness in the environment of unknown statistical distribution. We focus on fully-distributed online learning by agents and employ penalized individual cost functions to deal with coupled constraints. 
Three stochastic gradient strategies are developed with constant step-sizes. We allow the agents to use heterogeneous step-sizes and show that the penalty solution is able to approach the Nash equilibrium in a stable manner within $O(\mu_\text{max})$, for small step-size value $\mu_\text{max}$ and sufficiently large penalty parameters. The operation of the algorithm is illustrated by considering the network Cournot competition problem.
\end{abstract}


\IEEEpeerreviewmaketitle

\section{Introduction}

\IEEEPARstart{T}{he} generalized Nash equilibrium problem (GNEP) refers to a setting where each agent in a collection of agents seeks to minimize its own cost function subject to certain constraints {\em and} where both the cost function and the constraints are generally dependent on the
actions selected by the other agents~\cite{Debreu52,Harker91,Contreras04,Pang05,Facchinei07,Fischer14}. The GNEP was first formally introduced in~\cite{Debreu52} and was called a social equilibrium problem. A special case of GNEPs was considered in the work~\cite{Rosen65} where all agents shared common constraints. GNEPs arise naturally in the modeling of many applications, ranging from market liberalization of electricity~\cite{Contreras04,Hobbs07}, to natural gas~\cite{Abada13}, telecommunications~\cite{Chung03}, femto-cell power allocation~\cite{Ghosh15}, environmental pollution control~\cite{Krawczyk05}, and cloud computing~\cite{Ardagna13,Cardellini15}. Useful overviews on GNEPs appear in~\cite{Facchinei07,Fischer14}. 

In these types of problems, the Nash equilibrium is a desired and stable solution since at the Nash equilibrium no agent can benefit by unilaterally deviating from the solution. However, Nash equilibrium solutions may not exist or may not be unique. For instance, it was shown in~\cite{Harker91,Pang05} that the solution set of a GNEP can be characterized by solving a quasi-variational inequality (QVI), and it is rare that explicit results in QVIs can be utilized in GNEPs. Still, there is one common and important class of GNEPs that can be partially solved by solving a variational inequality (VI)~\cite{Facchinei07,Scutari10}. In this work, we focus on GNEPs with shared and coupled constraints since the theory of variational inequalities (VI) is more mature and has more useful results than the theory of quasi-variational inequalities (QVI).

In general, GNEP formulations do not admit closed-form solutions and many algorithms have been proposed to compute the solutions numerically. For example, GNEPs can be reformulated and solved using Nikaido-Isoda (NI) functions. Minimizing the NI can be achieved by means of gradient-descent algorithms~\cite{Heusinger09} or relaxtion-based algorithms~\cite{Heusinger09b}. Likewise, using the Karush–Kuhn–Tucker (KKT) conditions, GNEPs can be solved numerically, as demonstrated in~\cite{Facchinei14}. One can also resort to penalty-based reformulations where the original cost function is modified by including a penalty term. The purpose of the penalty term is to assign large penalties to deviations from the constraints. The works in~\cite{Facchinei10,Fukushima11} consider exact penalty functions and focus on updating the penalty parameters incrementally until a certain stopping rule is satisfied. 

In all these prior works~\cite{Debreu52}--\hspace{-0.1mm}\cite{Fukushima11}, the individual cost functions are assumed to be {\em deterministic}. This means that, when seeking GNEP solutions, we are able to acquire exactly the NI functions or the gradient vectors as necessary. However, when the agents are subject to randomness in the environment, it is customary to define the cost functions in terms of {\em expectations} of certain loss functions. The expectation operations are in relation to the distribution of the random data, which is rarely known beforehand. This {\em stochastic} type of Nash games arises in many practical applications, e.g., in the transportation model of~\cite{Watling06} and the signal transmission model for wireless networks in~\cite{Ghosh15}. To deal with stochasticity, the sample average approximate (SAA) method was proposed in\cite{Xu13} to approximate the expectation of the individual cost functions. However, in this method, the equilibrium solutions are learned in an off-line manner and the GNEP needs to be re-solved for every given batch of samples. 

\begin{table*}[t!] 
	\caption{Comparing with Existing Works for Distributed Stochastic Problems.} 
	\footnotesize
	\centering 
	\begin{tabu}{|[0.5pt]c||c|c|c|c|c|c|c|c|[0.5pt]}
		\thickhline
		\cellcolor[HTML]{ECF4FF} 
		& \cellcolor[HTML]{ECF4FF} Optimization Target 
		& \cellcolor[HTML]{ECF4FF} Constraints 
		& \cellcolor[HTML]{ECF4FF} Feasibility Approach
		& \cellcolor[HTML]{ECF4FF} Step-Sizes  
		& \cellcolor[HTML]{ECF4FF} Iterates Feasibility
		& \cellcolor[HTML]{ECF4FF} Tracking Ability \\
		\hline \hline 
		\cellcolor[HTML]{ECF4FF} Regularized SA\cite{Koshal13}
		& \tabincell{c}{Monotone \\ individual cost}
		& \tabincell{c}{Shared and \\ coupled}
		& Uses projection 
		& \tabincell{c}{Heterogeneous \\ decaying} 
		& Feasible 
		& No \\ 
		\hline
		\cellcolor[HTML]{ECF4FF} Penalized Diffusion \cite{Towfic14}
		& \tabincell{c}{Strongly-convex \\ aggregate cost}
		& Decoupled 
		& Uses penalty functions 
		& \tabincell{c}{Uniform \\ constant}
		& \tabincell{c}{Asymptotically \\ feasible}  
		& Yes \\ 
		\hline		
		\cellcolor[HTML]{ECF4FF} This Work 
		& \tabincell{c}{Strongly-monotone \\ individual costs}
		& \tabincell{c}{Shared and \\ coupled}
		& Uses penalty functions		
		& \tabincell{c}{Heterogeneous \\ constant} 
		& \tabincell{c}{Asymptotically \\ feasible}
		& Yes \\ 
		\thickhline 
	\end{tabu}
	\label{table1}
\end{table*}

In order to attain continuous learning in an online manner, the stochastic approximation (SA) method is a more suitable approach for differentiable cost functions, where the true gradient vectors are replaced by approximations. One stochastic implementation along these lines is considered in~\cite{Koshal13} albeit with a {\em vanishing} step-size parameter. The use of step-sizes that decay to zero is problematic in scenarios that require continuous adaptation and learning. 

For example, in nonstationary environments, the Nash equilibrium will drift with time due to changes in the statistical distribution of subsequent changes in the locations of the minimizers of the cost functions. 
When the step-size approaches zero, as is the case with the rules considered in~\cite{Bubeck12,Hazan12,Shalev11,Zinkevich03}, adaptation stops and the stochastic gradient algorithm loses its ability to track the drift. The approach in~\cite{Besbes15} employs a decaying step-size to track the evolving minimizer of a non-stationary objective. However, in that work, the optimal sublinear regret is obtained under the condition that the variation budget $V_T$ of the time-varying loss functions is sublinear with time. This condition implies that the variation in the loss functions should diminish with time, which is not applicable in the case where the minimizer of the cost function drifts continuously. One example that does not satisfy the variation budget condition is discussed in~\cite{Towfic14}. 
In comparison, it is well-known that constant step-size adaptation in inherently capable of tracking moderate drifts due to nonstationarity in the data --- see, e.g., the analysis in~\cite{Polyak87,Sutton17,Sayed142}.
 
We therefore focus in this work on online and fully-distributed learning to solve the stochastic GNEPs where agents are only allowed to interact locally with their neighbors. 
We assume that such interactions are confined to neighboring agents over the network topology and are subject to some coupled constraints shared by all neighbors. That is, in addition to the stochastic setting, we build one additional topology layer on top of conventional GNEPs with shared constraints. One example for such stochastic GNEP scenarios linked to a geometric topology would be the femto-cell power allocation problem considered in~\cite{Ghosh15}, where distributed algorithms are proposed and designed for this specific application. In this work, we study general distributed learning strategies for the solution of GNEPs by networked agents. 
Motivated by results from~\cite{Facchinei10,Fukushima11,Towfic14,Pinar94}, we first resort to penalty functions to deal with the constraints in stochastic GNEPs. The penalty reformulation helps avoid the high computational complexity of conventional NI-based approaches or the requirement of projection steps.
Traditionally, penalty methods focus on selecting penalty parameters~\cite{Facchinei10,Fukushima11}. However, in order to cope with the stochastic nature of GNEPs, we fix the penalty parameters at constant but sufficiently large values, in a manner similar to~\cite{Towfic14,Pinar94}, and study the resulting performance under stochastic environments.
We also focus on the use of {\em constant} step-sizes in the stochastic approximation  methods to enable continuous adaptation and learning. When this is done, gradient noise seeps into the operation of the algorithm. By gradient noise we mean the difference between the true gradient vector and its approximation. In decaying step-size implementations, this gradient noise component is annihilated over time by the diminishing step-size parameter at the expense of a deteriorating tracking performance. In contrast, in the constant step-size implementation, the gradient noise process is persistently present in the operation of the algorithm. One main challenge in our analysis is to establish that the stochastic-gradient implementation is able to keep the influence of gradient noise under check and to deliver an accurate estimation of the Nash equilibrium. Arriving at these conclusions for networked agents is one key contribution of this work. In Table~\ref{table1} we list a summary of properties comparing our results to two other existing works for distributed stochastic problems.

We remark that there exist other techniques in the stochastic optimization literature to solve problems with the variational inequalities. For example, the works \cite{Nesterov09,Xiao10,Duchi12} consider a dual-averaging method, which requires the solution of an optimization problem at each iteration; this formulation would be useful in situations when the optimization problem can be solved in closed form. References~\cite{Juditsky11,Mertikopoulos16} consider stochastic mirror-based approaches, which assume the gradient noise has bounded variance. It is worth noting that the methods in these earlier references are not directly applicable to GNEP with shared constraints over networks, which is one  critical contribution in this article. 

In the simulations section, we will illustrate the theoretical results and apply the proposed algorithms to the constrained network Cournot competition problem, which is widely used in applications such as economic trading with geographical considerations, power management over smart grids, and resource allocation~\cite{Contreras04,Hobbs07,Bimpikis14,Parzy10}. We will assume there that factories and markets are connected in a Cournot network and suffer from some randomness in the parameters. We will see that the numerical results will match well with our theoretical analysis. We will also compare our algorithms with two projection-based algorithms from~\cite{Koshal13} with decaying step-sizes: the distributed Arrow-Hurwicz method and the iterative Tikhonov regularization method. We will find that our algorithms converge faster; while the mean-square-error of the method with decaying step-sizes continuously improves at the expense of loss in tracking and adaptation abilities. 

\textbf{Notation}: We use lowercase letters to denote vectors and scalars, uppercase letters for matrices, plain letters for deterministic variables, and boldface letters for random variables. Table~\ref{table:symbol} provides a summary of the symbols used in the article for ease of reference.

\begin{table}[t!]
	\caption{Summary of Main Symbols and Notation.} 
	\footnotesize
	\centering
		\begin{tabu}{|c||l|c|l|}
			\thickhline
			\rowfont[c]  
			\footnotesize Symbol \cellcolor[HTML]{ECF4FF}
			& Meaning \cellcolor[HTML]{ECF4FF} 
			& Equation \cellcolor[HTML]{ECF4FF} 
			\\ 
			\hline \hline 
			$J_k(\cdot), Q_k(\cdot)$ 
			& Individual cost and loss functions
			& (\ref{opt_constraint}) \\ 
			\hline
			$J_k^p(\cdot)$ 
			& Penalized individual cost function
			& (\ref{opt_reform}) \\ 
			\hline
			$p_k(\cdot)$ & Aggregated penalty function
			& (\ref{pk})  \\
			\hline
			$F(w)$	
			& Block gradient vector
			& (\ref{blockgrad})	\\ 
			\hline
			$F^p(w)$	
			& Penalized block gradient vector
			& (\ref{penaltyblockgrad})	\\ 
			\hline
			$\nu$
			& Strongly-monotone parameter
			& (\ref{strongmono}) \\
			\hline
			$\delta$ 
			& Lipschitz parameter
			& (\ref{Lipschitz}) \\
			\hline
			$\rho$ 
			& Penalty parameter
			& (\ref{opt_reform})  \\
			\hline
			$\gamma_k$ 
			& Lipschitz gradient parameter 
			& (\ref{penalLipschitz})  \\
			\hline
			$\delta_p$ 
			& Parameter related to $\gamma_k$
			& (\ref{combineLip})  \\
			\hline
			$\mu_\text{max}$ 
			& Maximal step-size
			& (\ref{maxmu})  \\
			\hline
			$t$ 
			& Difference parameter for step-sizes
			& (\ref{minmu})  \\
			\hline
			$\alpha$ 
			& Gradient noise parameter
			& (\ref{gradnoise2})  \\
			\hline
			$\nu'$, $\nu''$ 
			& Weighted strongly-monotone parameters
			& (\ref{heterstrongmono}), (\ref{heterblock})  \\
			\hline
			\thickhline
		\end{tabu}
	\label{table:symbol}
\end{table}

\section{Problem Setup}

\begin{figure}[t!]	
	\begin{minipage}[b]{1.0\linewidth}
		\centering
		\centerline{\includegraphics[width=1in]{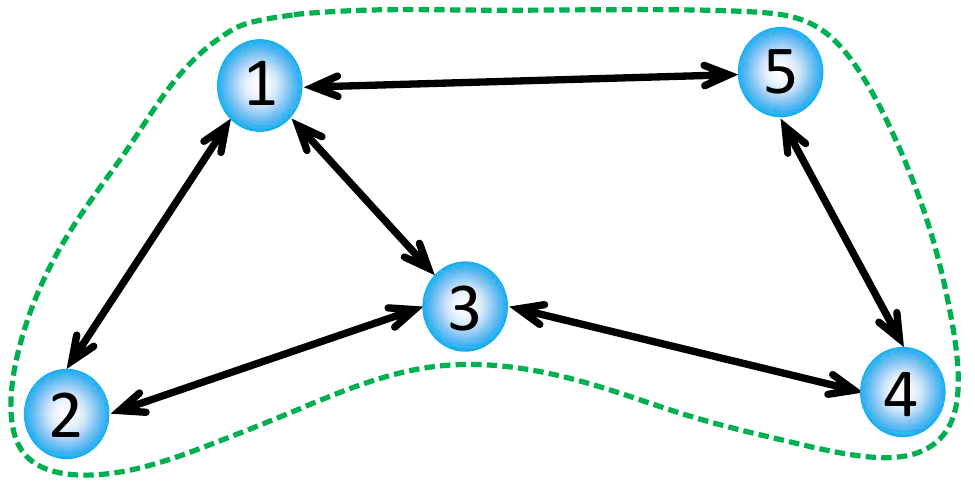}}
	\end{minipage}
	\caption{\small Illustration of the shared constraints over a network topology where $\mathds{1}$ denotes the vector with all one entries.}
	\label{topology}
\end{figure}

Consider a connected network of $N$ agents indexed by the set $\mathcal{N} = \{1,...,N\}$. 
The neighborhood of each agent $k$, denoted by $\mathcal{N}_k$, includes agent $k$ and the neighboring agents connected to $k$. We denote the action of each agent $k$ by a vector $w_k \in \mathbb{R}^{M_k}$ and associate with $k$ an individual risk function denoted by $J_k(\cdot)$. The argument of $J_k(\cdot)$ does not depend solely on $w_k$ but also on the action vectors of the neighboring agents. 
We collect the actions of all agents in the neighborhood $\mathcal{N}_k$ into the block vector:
\begin{align}
w^k
= \text{col}\{w_\ell; \ell \in \mathcal{N}_k\} \in \mathbb{R}^{M^k}
\end{align}
and the actions of all agents in the network $\mathcal{N}$ into:
\begin{align}
w = \text{col}\{w_1,\dots,w_N\} \in \mathbb{R}^M
\end{align}
where 
\begin{align}
M^k \triangleq \sum_{\ell \in \mathcal{N}_k} M_\ell, \quad M \triangleq \sum_{\ell=1}^N M_\ell
\end{align}
For convenience, we also introduce the notation
\begin{align}
w_{-k} \triangleq \text{col} \{w_\ell; \ell \in \mathcal{N}_k \setminus \{k\} \}
\end{align}
to collect the actions of all other agents in $\mathcal{N}_k$, with the exception of agent $k$. Using this notation, we shall sometimes write $J_k(w_k;w_{-k})$ instead of $J_k(w^k)$ in order to make the dependence on $w_k$ explicit. 
We consider that the action of each agent $k$ should satisfy a set of local constraints: 
\begin{align}
	h_{k,u}(w^k) &= 0, \quad u=1,\dots,U_k, \\
	g_{k,q}(w^k) &\leq 0, \quad q=1, \dots, L_k
\end{align}
The local constraint functions $\{h_{k,u}(w^k), g_{k,q}(w^k)\}$ at agent $k$ are assumed to be differentiable and known to agent $k$. We also assume that the equality constraint functions $\{h_{k,u}(w^k)\}$ are affine and the inequality functions $\{g_{k,q}(w^k)\}$ are convex in $w^k$. We further assume that the constraints are shared by the neighbors, i.e., if the argument of any $h_{k,u}(w^k)$ or $g_{k,q}(w^k)$ at node $k$ contains the action of some neighbor $\ell\in \mathcal{N}_k$, then agent $\ell$ is subject to the same constraint function, i.e., it will hold that $h_{\ell,u'}(w^\ell)=h_{k,u}(w^k)$ or $g_{\ell,q'}(w^\ell)=g_{k,q}(w^k)$ for some $u'$ and $q'$. Figure~\ref{topology} illustrates this setting for a network topology with $5$ agents.
An example of shared constraints is $g_{1,1}(w^1)=g_{2,1}(w^2)=g_{3,1}(w^3) \leq 0$, which is shared by the connected agents $1,2$ and $3$. We note that while there is no direct link between agents $2$ and $4$, the actions for these agents are coupled through the intermediate agent $3$. Therefore, in general, the actions of agents are affected explicitly by the neighbors and also implicitly by other agents in the network. This scenario is common in applications~\cite{Fukushima11,Facchinei07,Rosen65,Scutari10}. Each agent $k$ then seeks an optimal action vector that solves the following constrained optimization problem~\cite{Ravat11,Koshal13,Xu13}:
\begin{align}
\label{opt_constraint}
\min\limits_{w_k\in \mathbb{R}^{M_k}} \quad~ & \quad J_k (w^k) \triangleq \mathbb{E}_{\bm{x}_k} Q_k(w^k;\bm{x}_k) \notag \\
\text{subject to} & \quad
h_{k,u}(w^k) = 0, \quad u=1,\dots,U_k \notag \\
& \quad g_{k,q}(w^k) \leq 0, \quad q=1, \dots, L_k
\end{align} 
where $J_k (w^k)$ is assumed to be differentiable and strongly-convex in $w_k$, $Q_k(\cdot)$ is a scalar-valued loss function for agent $k$, and the expectation is taken over the distribution of the random data $\bm{x_k}$. 
For example, if we consider power allocation in wireless heterogeneous networks, the individual cost function $J_k(w^k)$ for each femto-base station $k$ can represent the Shannon capacity function with channel uncertainty. Moreover, one constraint of $g_{k,q}(w^k)$ shared by neighboring femto-base stations can be used to guarantee that the average signal-to-interference and noise ratio (SINR) at macro-user terminals is above a certain threshold~\cite{Ghosh15}. Problem (\ref{opt_constraint}) is known as the stochastic generalized Nash equilibrium problem (GNEP). 
For convenience, we collect all {\em distinct} individual constraints across all agents into a global set denoted by 
\begin{align}
\label{distinctpenalty}
\mathcal{S} \triangleq \{w; h_{u}(w) = 0, g_{q}(w) \leq 0, 1\leq u \leq U, 1\leq q \leq L\}
\end{align}
by removing the repeated shared constraints. 
We assume that ${\cal S}$ is nonempty, which means that at least one solution $w$ exists that satisfies the constraints in ${\cal S}$ and implies that the GNEP in (\ref{opt_constraint}) is feasible for each agent.
Let us denote the feasible set of (\ref{opt_constraint}) by 
\begin{align}
\mathcal{S}_k(w_{-k}) &\triangleq \{w_k ; h_{k,u}(w^k) = 0,   g_{k,q}(w^k) \leq 0, \notag \\
&\qquad \qquad \qquad 1\leq u \leq U_k, 1\leq q \leq L_k\}
\end{align}
Without loss of generality, we assume that the input (domain) of $\mathcal{S}_k(w_{-k})$ satisfies all constraints in $\mathcal{S}$ that are independent of $w_k$.
Therefore, any $w_k \in \mathcal{S}_k(w^{-k})$ shall satisfy the remaining constraints in $\mathcal{S}$ that are related to $w_k$, i.e., for each agent $k$ we have
\begin{align}
\label{sharedconst}
\mathcal{S}_k(w_{-k}) = \mathcal{S}_k (w^{-k}) &= \{w_k; (w_k,w^{-k})\in \mathcal{S}\} 
\end{align}
where
\begin{align}
w^{-k} \triangleq \text{col} \{w_\ell; \ell \in \mathcal{N} \setminus \{k\} \}
\end{align}
since the actions of the agents who are not neighbors of agent $k$ will not appear in any argument of the constraint functions $h_{k,u}(w^k)$ and $g_{k,q}(w^k)$. The conclusion in (\ref{sharedconst}) shows that the scenario considered in this work satisfies the condition of GNEP with general shared common constraints~\cite{Facchinei07}.

Our objective now is to derive distributed learning strategies by which agents can adaptively learn to solve (\ref{opt_constraint}) using local observations of the actions of neighboring agents.
\begin{table}[t!]
	\large
	\resizebox{\columnwidth}{!}{
		\tabulinesep=0.8mm
		\begin{tabu}{|c||l|l|l|}
			\tthickhline
			\rowfont[c]  
			\Large Agent \cellcolor[HTML]{ECF4FF}
			& Neighborhood \cellcolor[HTML]{ECF4FF}
			& Individual Cost \cellcolor[HTML]{ECF4FF}  
			& Constraints \cellcolor[HTML]{ECF4FF} \\ 
			\hline \hline 
			\!\cellcolor[HTML]{ECF4FF}
			1
			& $\mathcal{N}_1=\{1, 2, 3, 5\}$ 
			& $J_1 (w^1) = \|w_1+w_5\|^2$ 
			& $g_{1,1}(w^1)=\|w_1\|^2 +\|w_2+w_3\|^2 -2 \leq 0$	\\ 
			\hline
			\!\cellcolor[HTML]{ECF4FF} 2
			& $\mathcal{N}_2=\{1, 2, 3\}$ 
			& $J_2 (w^2) = \|w_2\|^2$ 
			& \tabincell{l}{$g_{2,1}(w^2)=\|w_1\|^2 +\|w_2+w_3\|^2 -2 \leq 0$ \\$g_{2,2}(w^2)=\|w_2-w_3\|^2 -5 \leq 0$}	\\ 
			\hline
			\!\cellcolor[HTML]{ECF4FF} 3
			& $\mathcal{N}_3=\{1, 2, 3, 4\}$ 
			& $J_3 (w^3) = \|w_3\| \cdot \|w_4\|^2$ 
			& \tabincell{l}{$g_{3,1}(w^3)=\|w_1\|^2 +\|w_2+w_3\|^2 -2 \leq 0$ \\$g_{3,2}(w^3)=\|w_2-w_3\|^2 -5 \leq 0$}	\\ 
			\hline
			\!\cellcolor[HTML]{ECF4FF} 4
			& $\mathcal{N}_4=\{3, 4, 5\}$ 
			& $J_4 (w^4) = \|w_3\|+\|w_4\|^2$ 
			& $h_{4,1}(w^4)=\mathds{1}_{M_4}^{\sf T} w_4 + \mathds{1}_{M_5}^{\sf T} w_5 -1=0$ \\
			\hline
			\!\cellcolor[HTML]{ECF4FF} 5
			& $\mathcal{N}_5=\{1, 4, 5\}$ 
			& $J_5 (w^5) = \|w_1\| \cdot \|w_5\|^2$ 
			& $h_{5,1}(w^5)=\mathds{1}_{M_4}^{\sf T} w_4 + \mathds{1}_{M_5}^{\sf T} w_5 -1=0$ \\
			\tthickhline
		\end{tabu}
	}
\end{table}

In preparation for our development, we collect the individual gradient vectors of $\{J_k(w^k)\}$ with respect to each $w_k^{\sf T}$ into 
\begin{align}
\label{blockgrad}
F(w) \triangleq \text{col}\{\nabla_{w_1^{\sf T}} J_1(w^1),...,\nabla_{w_N^{\sf T}} J_N(w^N)\}
\end{align}
and assume that this block column vector satisfies the following properties.
\begin{asump} ({\bf$\nu$-Strongly Monotone}) \label{assump:strongmono}
	For any two action profiles $w=w^\circ$ and $w=w^\bullet$, it holds that
	\begin{align}
	\label{strongmono}
	\left(w^\circ - w^\bullet\right)^{\sf T} [F(w^\circ) - F(w^\bullet)]
	&\geq \nu \|w^\circ - w^\bullet\|^2
	\end{align}
	for some positive constant $\nu$. \QEDB
\end{asump}
\begin{asump} ({\bf$\delta$-Lipschitz Continuous}) \label{assump:Lipschitz}
	The block column vector $F(w)$ is assumed to be Lipschitz continuous, i.e.,
	\begin{align}
	\label{Lipschitz}
	\|F(w^\circ) - F(w^\bullet)\| \leq \delta \|w^\circ-w^\bullet\|
	\end{align}
	for some positive constant $\delta$. \QEDB
\end{asump}
If we consider two action vectors $w^a$ and $w^b$ defined as:
\begin{align}
w^a &\triangleq \text{col}\{w_1,...,w_k^\circ,...,w_N\} \\
w^b &\triangleq \text{col}\{w_1,...,w_k^\bullet,...,w_N\}
\end{align}
for some $k$, then using (\ref{strongmono}) we get
\begin{align}
\label{stronglyconvex}
(w^a &- w^b)^{\sf T} [F(w^a) - F(w^b)] \notag \\
&= (w_k^\circ - w_k^\bullet)^{\sf T} \left[\nabla_{w_k^{\sf T}} J_k(w_k^\circ;w_{-k}) - \nabla_{w_k^{\sf T}} J_k(w_k^\bullet;w_{-k})\right] \notag \\
&\geq \nu \|w_k^\circ - w_k^\bullet\|^2
\end{align}
Therefore, Assumption~\ref{assump:strongmono} implies that each individual cost function $J_k(w^k)$ is strongly convex in $w_k$. 
Moreover, it holds that $\delta \geq \nu$ since from the Cauchy-Schwarz inequality we have
\begin{align}
\label{vdelta}
\nu \|w^\circ - w^\bullet\|^2 &\leq (w^\circ - w^\bullet)^{\sf T} [F(w^\circ) - F(w^\bullet)] \notag \\
&\leq \|w^\circ-w^\bullet\| \cdot \|F(w^\circ) - F(w^\bullet)\| \notag \\
&\leq \delta \|w^\circ-w^\bullet\|^2
\end{align}
\begin{examp} ({\bf Quadratic Risks})
One useful example of a loss function is the quadratic loss, which can be expressed in the following form with the entries of $\bm{x}_k$ split into $\bm{x}_k \triangleq \{\bm{B}_k, \bm{b}_k, \bm{\varepsilon}_k\}$:
\begin{align}
\label{IndivCost}
Q_k(w^k;\bm{x}_k) &= w^{k \sf T} \bm{B}_k w^k + \bm{b}_k^{\sf T} w^k + \bm{\varepsilon}_k \notag \\
&=\sum_{s \in \mathcal{N}_k} \sum_{\ell \in \mathcal{N}_k} w_s^{\sf T} \bm{B}_{s \ell}^k w_\ell + \sum_{\ell \in \mathcal{N}_k} \bm{b}_{k\ell}^{\sf T} w_\ell + \bm{\varepsilon}_k 
\end{align}
where $\bm{B}_k$ is a random symmetric matrix of size $M^k \times M^k$, $\bm{b}_k$ is a random vector of size $1 \times M^k$, and $\bm{\varepsilon}_k$ is a random scalar variable with mean $\varepsilon_k$. In (\ref{IndivCost}), we partitioned $\bm{B}_k$ and $\bm{b}_k$, respectively, into block matrices $\{\bm{B}_{s \ell}^k \in \mathbb{R}^{M_s \times M_\ell}\}$ and block vectors $\{\bm{b}_{k\ell} \in \mathbb{R}^{M_\ell \times 1}\}$ in conformity with the block structure of $w^k$. 
The random data $\{\bm{B}_k, \bm{b}_k, \bm{\varepsilon}_k\}$ are assumed to be independent of each other. Note that under (\ref{IndivCost}), the gradient vector of $J_k(w^k)$ with respect to $w_k^{\sf T}$ is the $M_k \times 1$ vector given by
\begin{align}
\nabla_{w_k^{\sf T}} J_k(w^k) 
&= \sum_{\ell \in \mathcal{N}_k} 2 B_{k \ell}^k w_{\ell} + b_{kk}
\end{align}
where we introduced the means $B_{k \ell}^k = \mathbb{E} \bm{B}_{k \ell}^k$ and $b_{k k} = \mathbb{E}\bm{b}_{k k}$. 
Collecting these individual gradient vectors we get 
\begin{align}
\label{Fw}
F(w) = B w + b
\end{align}
where 
\begin{align}
B \triangleq \begin{bmatrix}
2B_{11}^1 & \cdots & 2B_{1N}^1\\
\vdots & \ddots & \vdots \\
2B_{N1}^N & \cdots & 2B_{NN}^N
\end{bmatrix} \in \mathbb{R}^{M \times M}, b \triangleq \begin{bmatrix}
b_{11}\\
\vdots \\
b_{NN}
\end{bmatrix} \in \mathbb{R}^{M \times 1}
\end{align}
Note that Assumption~\ref{assump:strongmono} will hold if there exists a positive constant $\nu$ such that for any $M \times 1$ vector $a$ we have
\begin{align}
\label{SemiPosDef}
a^{\sf T} \left(B-\nu I\right) a \geq 0 \quad \Longleftrightarrow &\quad a^{\sf T} B a \geq \nu \|a\|^2 
\end{align}	
Since $B$ is not necessarily symmetric, we know from~\cite[p. 259]{Johnson70} that (\ref{SemiPosDef}) holds if, and only if, the symmetric part of $B$ satisfies:
\begin{align}
\frac{1}{2} (B+B^{\sf T}) \geq \nu I
\end{align}
It follows from this condition that the largest singular value of $B$, denoted by $\sigma_\text{max}$, should be greater than or equal to $\nu$ since
\begin{align}
\label{sigma_nu}
\sigma_\text{max} = \|B\| \geq \Big\| \frac{1}{2} (B+B^{\sf T})\Big\| \geq \nu
\end{align}
From (\ref{Fw}), it is easy to verify that  Assumption~\ref{assump:Lipschitz} always holds for the quadratic loss function since
\begin{align}
\|F(w^\circ) - F(w^\bullet)\| &= \| B (w^\circ-w^\bullet) \| \notag \\
&\leq \sigma_\text{max} \|w^\circ-w^\bullet\|
\end{align}
\QEDB
\end{examp}

\section{Stochastic Penalty-Based Learning}

\subsection{Penalty Approximation for Coupled Constraints}

Solving the constrained optimization problem (\ref{opt_constraint}) is generally demanding and may not admit a closed-form solution. In this work, we resort to a penalty-based approach to replace the original problem by an unconstrained optimization problem and then show that the solution to the penalized problem tends asymptotically with the penalty parameter to the desired solution to (\ref{opt_constraint}). Even more importantly, we will show that the penalty-based approach enables the agents to employ {\em adaptive} learning strategies, which instantaneously approximate the unknown random individual cost functions and endow the agents with the ability to track variations in the location of the Nash equilibrium due to changes that may occur in the constraint conditions or cost measures.

The main motivation for penalty methods is to assign a large penalty weight whenever constraints are violated and a smaller or zero weight when the constraints are satisfied~\cite{Polyak87,Bazaraa93,Towfic14,Pang05}. More specifically, problem
(\ref{opt_constraint}) is replaced by the following unconstrained formulation:
\begin{align}
\label{opt_reform}
\min_{w_k\in \mathbb{R}^{M_k}} \quad  J_k (w^k) + \rho p_k(w^k) \triangleq J_k^p(w^k) = J_k^p(w_k;w_{-k})
\end{align}
where $\rho\geq 0$ is a penalty parameter, $p_k(w^k)$ denotes the penalty function for agent $k$ and is assumed to be of the following aggregate form, with one penalty factor applied to each constraint:
\begin{align}
\label{pk}
p_k(w^k) = \sum_{u=1}^{U_k}  \theta_\text{EP} \left(h_{k,u}(w^k)\right) + \sum_{q=1}^{L_k}  \theta_\text{IP} \left(g_{k,q}(w^k)\right)
\end{align}
where $\theta_\text{EP}(x)$ and $\theta_\text{IP}(x)$ are convex functions. The equality penalty factor $\theta_\text{EP}(x)$ returns zero value if the constraint is satisfied, i.e., when $h_{k,u}(w^k) = 0$, and introduces a large positive penalty if the constraint is violated, i.e., when $h_{k,u}(w^k) \neq 0$. For example, a continuous and differentiable choice for the equality penalty is the quadratic function:
\begin{align}
\label{penaltyEP}
\theta_\text{EP} (x)= x^2 
\end{align}
Since $h_{k,u}(w^k)$ is affine, a convex choice of $\theta_\text{EP}(\cdot)$ ensures the convexity of the function composition $\theta_\text{EP} \left(h_{k,u}(w^k)\right)$.
Similarly, the inequality penalty function $\theta_\text{IP}(x)$ returns zero value if $g_{k,q}(w^k) \leq 0$, and introduces a large positive penalty if $g_{k,q}(w^k)>0$. In the penalty method studied in~\cite{Fukushima11}, we get an exact Nash equilibrium solution to (\ref{opt_constraint}) as long as $\rho$ is sufficiently large and we use the $\ell_1$ penalty function~\cite{Bertsekas75}:
\begin{align}
	\label{optpenal}
	\theta_\text{IP}^e(x) = \max\{0,x\}
\end{align}
However, using this penalty function makes the objective function in (\ref{opt_reform}) non-differentiable, which limits the use of gradient-based adaptation rules~\cite{Nesterov04}. To avoid this difficulty, we can employ the following half-quadratic penalty function~\cite{Bazaraa93}, which is continuous, convex, nondecreasing, and once-differentiable:
\begin{align}
	\label{penalty}
	\theta_\text{IP}(x) \triangleq \begin{cases}
	0, & x\leq 0 \\
	x^2/2, & x\geq 0 
	\end{cases}
\end{align}
Other choices for $\theta_\text{IP}(x)$ are of course possible, e.g., $\gamma$-norm\cite{Facchinei10}, exponential and shifted logarithmic functions~\cite{Ben97,Tseng93}, linear-quadratic functions~\cite{Pinar94}, and others in~\cite{Towfic14,Sayed14}. We note that a convex and nondecreasing choice of $\theta_\text{IP}(\cdot)$ results in a convex composite function $\theta_\text{IP} \left(g_{k,q}(w^k)\right)$ since $g_{k,q}(w^k)$ is convex. Consequently, the penalty function $p_k(w^k)$ defined in (\ref{pk}) is convex in $w^k$.

The penalized cost $J_k^p(w^k)$ is strongly-convex in $w_k$ since $J_k(w^k)$ is strongly-convex in $w_k$, as seen in (\ref{stronglyconvex}), and $p_k(w^k)$ is convex in $w^k$, and therefore in $w_k$.
An action profile $w^\star = \text{col}\{w^\star_1,...,w^\star_N\}$ that minimizes simultaneously all penalized costs $\{J_k^p(w^k)\}$ is called a Nash equilibrium for the penalized formulation (\ref{opt_reform}), i.e., for each agent $k$, the Nash equilibrium $w^\star$ satisfies
\begin{align}
J_k^p(w^\star_k;w^\star_{-k}) \leq J_k^p(w_k;w^\star_{-k}), \qquad  \forall w_k \in \mathbb{R}^{M_k}
\end{align}
The following theorem ensures the existence and uniqueness of the Nash equilibrium.
\begin{thm} ({\bf Existence and Uniqueness}): \label{UNIQUENESS}
	Under Assumption~\ref{assump:strongmono} and for any convex choice of $\theta_\text{EP}(x)$ and any convex and nondecreasing choice of $\theta_\text{IP}(x)$, there exists a unique Nash equilibrium $w^\star$ for problem (\ref{opt_reform}), and it satisfies
	\begin{align}
	\label{NashCond}
	F^p(w^\star) \triangleq F(w^\star) + \rho \nabla_{w^{\sf T}} p(w^\star) =0
	\end{align}
	where 
	\begin{align}
	\label{penaltyblockgrad}
	F^p(w) &\triangleq \textup{ col}\{\nabla_{w_1^{\sf T}} J_1^p(w^1),...,\nabla_{w_N^{\sf T}} J_N^p(w^N)\} \\
	\label{aggpenal}
	\nabla_{w^{\sf T}} p(w) &\triangleq \textup{col}\{ \nabla_{w_1^{\sf T}} p_1(w^1),...,\nabla_{w_N^{\sf T}} p_N(w^N) \}
	\end{align}
\end{thm}
\begin{IEEEproof} 
	See Appendix~\ref{proof:uniqueness}.
\end{IEEEproof}
Now, for any $\rho$, let us denote the unique Nash equilibrium to the penalized optimization problem (\ref{opt_reform}) by
\begin{align}
w^\star(\rho) &\triangleq \text{col}\{w^\star_1(\rho),...,w^\star_N(\rho)\} 
\end{align}
where 
\begin{align}
w^\star_k(\rho) &= \arg \min_{w_k \in \mathbb{R}^{M_k}} 
J_k^p(w_k;w^\star_{-k}(\rho)) \notag \\
&= \arg \min_{w_k \in \mathbb{R}^{M_k}} J_k (w_k;w^\star_{-k}(\rho)) + \rho p_k(w_k;w^\star_{-k}(\rho))
\end{align}
For convenience, we introduce the notation:
\begin{align}
w_k^\star(\infty) &\triangleq \lim_{\rho \rightarrow \infty} w_k^\star(\rho)\\
w_{-k}^\star(\infty) &\triangleq \text{col} \{w_\ell^\star(\infty); \ell \in \mathcal{N}_k \setminus \{k\} \} 
\end{align}
From the results in~\cite[p. 3930]{Towfic14} and~\cite[Theorem 9.2.2]{Bazaraa93}, we know that given any $w_{-k}$ and as $\rho$ goes to infinity, we have 
\begin{align}
\label{zaidlimit}
\inf_{w_k\in \mathcal{S}_k(w_{-k})} J_k (w_k;w_{-k}) =
\lim_{\rho \rightarrow \infty} \inf_{w_k\in \mathbb{R}^{M_k}} J_k^p(w_k;w_{-k})
\end{align}
and 
\begin{align}
J_k (w_k^o;w_{-k}) &= 
\inf_{w_k \in \mathcal{S}_k(w_{-k})} J_k (w_k;w_{-k})
\end{align}
where $w_k^o \in \mathcal{S}_k (w_{-k})$ is feasible for optimization problem (\ref{opt_constraint}) and satisfies
\begin{align}
w_k^o &\triangleq \lim_{\rho \rightarrow \infty} \arg \min_{w_k \in \mathbb{R}^{M_k}} J_k^p(w_k;w_{-k})
\end{align}
Therefore, if we are given $w^\star_{-k}(\infty)$, we get 
\begin{align}
J_k (w^\star_k(\infty);w^\star_{-k}(\infty)) &=\inf_{w_k \in \mathcal{S}_k (w^\star_{-k}(\infty))} J_k (w_k;w^\star_{-k}(\infty))  \notag \\
&=
\lim_{\rho \rightarrow \infty} \inf_{w_k \in \mathbb{R}^{M_k}} J_k^p(w_k;w^\star_{-k}(\infty)) \notag \\
&=J_k^p(w^\star_k(\infty);w^\star_{-k}(\infty)) 
\end{align}
It then follows that $w^\star(\infty) \triangleq \text{col}\{w^\star_1(\infty),...,w^\star_N(\infty)\}$ is an asymptotic Nash equilibrium of GNEP in (\ref{opt_constraint}) as $\rho$ goes to infinity. Furthermore, for each agent $k$ the value of the original cost $J_k$ coincides with the value of the penalized cost $J_k^p$ at $w^\star(\infty)$. Consequently, the Nash equilibrium for the penalized problem (\ref{opt_reform}) can be made arbitrarily close to the set of Nash equilibria (if not unique) by choosing $\rho$ large enough. 
Comparing with the variational equilibrium concept discussed in~\cite{Facchinei07a}, the main difference here is that instead of solving an exact GNE directly, we introduce the differentiable penalty function $p(\cdot)$ to get an asymptotic solution, which is more practical computationally under stochastic environments as we will see in later sections.
	
\subsection{Stochastic Learning Dynamics}
\label{stochastic_learning}

The unknown statistical distribution of the data makes it impossible to solve the penalized optimization problem (\ref{opt_reform}) analytically. As a result, a closed form solution to problem (\ref{opt_reform}) is not generally possible. If this were possible, then the agents could learn $w^{\star}$ given knowledge of the other agents' actions; this solution method would lead to the best response dynamics~\cite{Matsui92}. Since this approach is rarely applicable, agents can instead appeal to learning strategies where they gradually approach the desired $w^{\star}$ through successive inference
from streaming data. For example, one well-known gradient-descent solution to update the agents' actions at discrete-time instants $i$ is to employ the following localized rule~\cite{Flam02,Shamma05,Na13}:  
\begin{align}
	\label{Gradient}
	w_{k,i} &= w_{k,i-1} - \mu_k \nabla_{w_k^{\sf T}} J_k^p (w_{i-1}^k) \notag \\
	&=w_{k,i-1} - \mu_k ( \nabla_{w_k^{\sf T}} J_k (w_{i-1}^k) + \rho \nabla_{w_k^{\sf T}} p_k(w_{i-1}^k)  ) 
\end{align}
where $\mu_k$ is the step-size for agent $k$. Alternatively, motivated by the arguments from~\cite{Towfic14}, one can implement (\ref{Gradient}) incrementally by using a two-step learning strategy to improve the individual costs and the penalty costs separately. For example, agent $k$ can use an Adapt-then-Penalize (ATP) diffusion learning strategy to update first the iterate along the negative gradient direction of the individual cost $J_k(\cdot)$ and then apply the correction along the gradient of the penalty term: 
\begin{numcases}{\text{(ATP)~}}
	\label{Gradient31} \psi_{k,i} = w_{k,i-1} - \mu_k \nabla_{w_k^{\sf T}} J_k (w_{i-1}^k) \hspace*{5em} \\
	\label{Gradient32} w_{k,i}=\psi_{k,i} - \mu_k \rho \nabla_{w_k^{\sf T}} p_k(\psi_i^k) \hspace*{5em}
\end{numcases}
where $\psi_{k,i}\in \mathbb{R}^{M_k}$ is an intermediate action of agent $k$ and, similar to $w_i^k$,  the notation $\psi_i^k$ collects the iterates $\psi_{\ell,i}$ from across the neighborhood of agent $k$. Agents can also switch the order of these two steps and use a Penalize-then-Adapt (PTA) diffusion learning strategy:
\begin{numcases}{\text{(PTA)~}}
	\label{Gradient41} \psi_{k,i} = w_{k,i-1} - \mu_k \rho \nabla_{w_k^{\sf T}} p_k(w_{i-1}^k) \hspace*{5em} \\
	\label{Gradient42} w_{k,i}=\psi_{k,i} - \mu_k \nabla_{w_k^{\sf T}} J_k (\psi_{i}^k) \hspace*{5em}
\end{numcases}
We note that in the gradient-based learning strategies of (\ref{Gradient}), (\ref{Gradient31})--(\ref{Gradient32}), and (\ref{Gradient41})--(\ref{Gradient42}), agents are assumed to be able to observe or acquire the intermediate actions taken by neighboring agents and then synchronously update their actions\footnote{We remark that asynchronous adaptation and learning is also possible, see~\cite{Zhao15,Nassif16} and the references therein.  We focus in this work on synchronous operation. }. Furthermore, when implementing these strategies, each agent $k$ requires knowledge of its own gradient quantities $\nabla_{w_k^{\sf T}} J_k (w^k)$ and $\nabla_{w_k^{\sf T}} p_k (w^k)$.
When the exact statistics of the data $\bm{x}_k$ are unavailable, we need to resort to instantaneous realizations $\{\bm{x}_{k,i}\}$ of these random variables at each time 
$i$ and estimate the gradient vectors by employing constructions based on the loss functions, i.e.,
\begin{align}
\widehat{\nabla_{w_k^{\sf T}}} J_k(w^k) &\triangleq 
\nabla_{w_k^{\sf T}} Q_k(w^k;\bm{x}_{k,i}) 
\end{align}
Using these estimates, we arrive at the following stochastic gradient implementation:
\begin{align}
\label{StoGradient}
\bm{w}_{k,i} =\bm{w}_{k,i-1} &- \mu_k \nabla_{w_k^{\sf T}} Q_k (\bm{w}_{i-1}^k;\bm{x}_{k,i}) \notag \\
&- \mu\rho \nabla_{w_k^{\sf T}} p_k(\bm{w}_{i-1}^k)
\end{align}
and the corresponding ATP and PTA diffusion versions:
\begin{numcases}{\!\left(\!\!\!\begin{array}{c}
\text{diffusion} \\
\text{ATP}
\end{array}\!\!\!\right)\!\!}
\!\!\label{StoGradient31} \bm{\psi}_{k,i}  =  \bm{w}_{k,i-1} - \mu_k \nabla_{w_k^{\sf T}} Q_k (\bm{w}_{i-1}^k;\bm{x}_{k,i}) \hspace*{2em} \\
\!\!\label{StoGradient32} \bm{w}_{k,i}  =  \bm{\psi}_{k,i} - \mu_k \rho \nabla_{w_k^{\sf T}} p_k(\bm{\psi}_i^k) \hspace*{2em}
\end{numcases}
and
\begin{numcases}{\!\!\left(\!\!\!\begin{array}{c}
	\text{diffusion} \\
	\text{PTA}
	\end{array}\!\!\!\right)\!\!}
\!\!\label{StoGradient41} \bm{\psi}_{k,i} = \bm{w}_{k,i-1} - \mu_k \rho \nabla_{w_k^{\sf T}} p_k(\bm{w}_{i-1}^k) \hspace*{4.5em} \\
\!\!\label{StoGradient42} \bm{w}_{k,i}=\bm{\psi}_{k,i} - \mu_k \nabla_{w_k^{\sf T}} Q_k (\bm{\psi}_i^k;\bm{x}_{k,i}) \hspace*{4.5em}
\end{numcases}
Observe that we are denoting the weight iterates in boldface since they are now random quantities due to the randomness of $\widehat{\nabla_{w_k^{\sf T}}} J_k(w^k)$ resulting from the use of realizations $\bm{x}_{k,i}$. Note also that instead of diminishing step-sizes, we are considering constant step-sizes $\{\mu_k\}$ in order to endow the algorithms with a tracking mechanism that enables them to track variations in the statistical distribution of the data over time. 
If the step-sizes are uniform, i.e., $\mu_k=\mu$ for all $k$, we will show later in Sec.~\ref{sec:analysis} that the diffusion ATP and PTA strategies are more stable than the stochastic gradient (\ref{StoGradient}). Furthermore, we will observe in the simulations of Sec.~\ref{cournotnet} that the diffusion ATP and PTA strategies exhibit better mean-square error performance than the stochastic gradient (\ref{StoGradient}).

\begin{examp} ({\bf Multitask Diffusion Adaptation})
The formulations of multitask diffusion adaptation in~\cite{Chen14,Chen15,Nassif16} can also be regarded as a special case of the GNEP formulation for quadratic cost functions. In multitask scenarios, there exist clusters in the network with agents in the same cluster interested in the same objective or task (such as estimating a common vector). Cooperation is still warranted among agents and clusters because the multiple tasks can have some similarities. We can reformulate the multitask problem as an GNEP as follows:
\begin{align}
\label{multitask}
\min\limits_{w_k} \quad~~~ & \quad ~ J_k (w^k)  \notag \\
\text{subject to} & \quad ~ w_k = w_\ell, ~ \ell \in \mathcal{N}_k \cap \mathcal{C}_k
\end{align} 	
where $\mathcal{C}_k$ denotes the cluster that agent $k$ belongs to. Note that the constraints are only on the neighboring agents belonging to cloud $\mathcal{C}_k$ since they have the same estimation target. Following~\cite{Chen14,Chen15,Nassif16}, we consider a regularized mean-square-error risk of the form:
\begin{align}
J_k (w^k) &\triangleq  \mathbb{E}|\bm{d}_k(i) - \bm{u}_{k,i} w_k|^2 +\!\!\! \sum_{\ell \in \mathcal{N}_k \setminus \mathcal{C}_k} \!\! \eta_{k\ell} \|w_k-w_\ell\|^2
\end{align} 
where the scalar $\bm{d}_k(i) \in \mathbb{R}$ and the regression vector $\bm{u}_{k,i} \in \mathbb{R}^{1\times M}$ are the observation data, and $\{\eta_{k \ell}\geq 0\}$ are regularization parameters. Note that the regularization terms include only the neighboring agents in different clusters from $\mathcal{C}_k$. 
Let us rewrite the constraints as $\{w_k(m) - w_\ell(m)=0\}$ for $\ell \in \mathcal{N}_k \cap \mathcal{C}_k$ and $m=1,...,M$, and then use the quadratic penalty function (\ref{penaltyEP}) to get
\begin{align}
\label{opt_reform_diff}
p_k(w^k) &= \sum_{\ell \in \mathcal{N}_k \cap \mathcal{C}_k} \sum_{m=1}^{M}   \left(w_k(m) - w_\ell(m) \right)^2
\end{align} 
with the gradient vector 
\begin{align}
\label{distopt}
\nabla_{w_k^{\sf T}} p_k(w^k) &= \sum_{\ell \in \mathcal{N}_k \cap \mathcal{C}_k} 2 (w_k - w_\ell) 
\end{align}
Using the diffusion ATP strategy in (\ref{StoGradient31})--(\ref{StoGradient32}), we then arrive at the multitask ATC algorithm derived in~\cite{Chen14,Chen15,Nassif16}:
\begin{numcases}{\!\!\!\left(\!\!\!\begin{array}{c}
	\text{multitask} \\
	\text{ATC}
	\end{array}\!\!\!\right)\!\!}
\label{multi_diff1} \bm{\psi}_{k,i}  =  \bm{w}_{k,i-1} + \mu_k \bm{u}_{k,i}^{\sf T} [\bm{d}_k(i) - \bm{u}_{k,i} \bm{w}_{k,i-1}]  \notag \\
\qquad\quad+\!\! \sum_{\ell \in \mathcal{N}_k \setminus \mathcal{C}_k} \eta_{k\ell} (\bm{w}_{\ell,i-1}-\bm{w}_{k,i-1}) \\
\label{multi_diff2} \bm{w}_{k,i}  =  \sum_{\ell \in \mathcal{N}_k \cap \mathcal{C}_k} a_{\ell k} \bm{\psi}_{\ell,i}
\end{numcases}
where 
\begin{align}
\label{combi_multi}
a_{k k} &\triangleq 1 - \sum_{\ell \in \mathcal{N}_k \cap \mathcal{C}_k} 2 \mu_k \rho, \quad
a_{\ell k} \triangleq 2 \mu_k \rho, \quad \text{for}~\ell \neq k
\end{align}
We note that for the case $\mathcal{C}_k=\mathcal{N}_k$, the multitask ATC algorithm (\ref{multi_diff1})--(\ref{multi_diff2}) becomes a standard diffusion strategy~\cite{Chen151,Chen152,Sayed14,Sayed132,Sayed142,Sayed143,Catt10,Chouvardas11,Chen12}. The consensus strategies~\cite{Braca08,Dimakis10,Olfati07,Kar11,Nedic09} can also be derived by considering the stochastic gradient descent rule (\ref{StoGradient}) and using similar arguments.
\QEDB
\end{examp}

\section{Performance Analysis}
\label{sec:analysis}

We now examine the convergence and stability properties of the distributed stochastic algorithms (\ref{StoGradient31})--(\ref{StoGradient32}) and (\ref{StoGradient41})--(\ref{StoGradient42}). In particular, we examine how close their limiting point gets to the unique equilibrium point, $w^{\star}$. To continue, we introduce the following condition on the penalty function. This condition is not restrictive since the choice of the penalty function is under the designer's control.
\begin{cond} ({\bf Lipschitz gradients})
\label{cond:LipschitzPw}
	Consider two arbitrary block vectors $w^\circ$ and $w^\bullet$ collecting all actions from  all agents: 
	\begin{align}
	w^\circ \triangleq \text{col} \{w^\circ_1,...,w^\circ_N\},\qquad 	w^\bullet \triangleq \text{col} \{w^\bullet_1,...,w^\bullet_N\}
	\end{align}
	We denote the corresponding action vectors in $\mathcal{N}_k$ by
	\begin{align}
	w^k_\circ \triangleq \text{col}\{w^\circ_\ell; \ell \in \mathcal{N}_k\},\qquad 	w^k_\bullet \triangleq \text{col}\{w^\bullet_\ell; \ell \in \mathcal{N}_k\}
	\end{align}
	For each individual agent $k$, we assume that the gradient vector $\nabla_{w_k^{\sf T}} p_k(\cdot)$ satisfies:
	\begin{align}
	\label{penalLipschitz}
	\left\| \nabla_{w_k^{\sf T}} p_k(w_\circ^k)-\nabla_{w_k^{\sf T}} p_k(w_\bullet^k)	\right\| \leq \gamma_k \left\|w_\circ^k-w_\bullet^k\right\|
	\end{align}
	where $\gamma_k$ is a positive constant.
	\QEDB
\end{cond}
Note that $p_k(w^k)$ is not required to be twice-differentiable, which is weaker than the assumption used in~\cite{Towfic14}.
Then, we have the following theorem.
\begin{lemma} ({\bf Lipschitz continuity}) \label{LIP_CONT}
	Under Condition~\ref{cond:LipschitzPw} and Assumption~\ref{assump:Lipschitz}, the penalized block gradient vector $F^p(w)$ is $(\delta+\rho\delta_p)$-Lipschitz continuous, i.e., for any $w^\circ$ and $w^\bullet$ we have
	\begin{align}
	\label{combineLip}
	\|F^p&(w^\circ) - F^p(w^\bullet)\| \leq (\delta+\rho \delta_p) \|w^\circ-w^\bullet\|
	\end{align}
	where $\delta_p \triangleq (\sum_{k=1}^{N} \gamma_k^2)^{1/2}$.
\end{lemma}
\begin{IEEEproof}
	See Appendix~\ref{proof:Lip_cont}.
\end{IEEEproof}
In order to characterize the heterogeneous step-sizes, let us denote the maximal and minimal step-sizes, respectively, over the network by
\begin{align}
\label{maxmu}
\mu_\text{max} &\triangleq \max_{1\leq k \leq N}\{\mu_k\}  \\
\label{minmu} 
\mu_\text{min} &\triangleq \min_{1\leq k \leq N}\{\mu_k\} \triangleq (1-t) \mu_\text{max}
\end{align}
for some parameter $0\leq t <1$. A small value of $t$ indicates that the step-sizes $\{\mu_k\}$ are clustered together. To continue, we establish the following lemma.
\begin{lemma} ({\bf Weighted strong monotonicity})
	\label{heterstrong}
	The penalized block gradient vector $F^p(w)$ satisfies, for any two action profiles $w^\circ$ and $w^\bullet$,  
	\begin{align}
	\label{heterstrongmono}
	(w^\circ - w^\bullet)^{\sf T} &U [F^p(w^\circ) - F^p(w^\bullet)] \geq \mu_\text{max}\nu' \|w^\circ - w^\bullet\|^2
	\end{align}
	where $U \triangleq \text{diag}\{\mu_1 I_{M_1},...,\mu_N I_{M_N}\}$ is a diagonal matrix with step-sizes in the diagonal positions and $\nu' \triangleq \nu-t(\delta+\rho \delta_p)$.
	Similarly, the block gradient vector $F(w)$ and the	
	penalty gradient vector $\nabla_{w^{\sf T}} p(w)$ satisfy, respectively,
	\begin{align}
	\label{heterblock}
	(w^\circ - w^\bullet)^{\sf T} U [F(w^\circ) - F(&w^\bullet)] \geq \mu_\text{max}\nu'' \|w^\circ - w^\bullet\|^2 \\
	\label{heterpenalty}
	(w^\circ - w^\bullet)^{\sf T} U [\nabla_{w^{\sf T}} p(w^\circ) - &\nabla_{w^{\sf T}} p(w^\bullet)] \notag \\
	&~\geq -t\mu_\text{max} \delta_p \|w^\circ - w^\bullet\|^2
	\end{align}
	where $\nu''\triangleq \nu-t\delta$.
\end{lemma}
\begin{IEEEproof}
	See Appendix~\ref{proof:heterstrong}.
\end{IEEEproof}
Note that for uniform step-sizes we have $t=0$ and thus $\nu'=\nu''=0$. Furthermore, $\nu'$ and $\nu''$ are not necessarily positive unless $t$ is small enough.
We further introduce the gradient noise vector
\begin{align}
\bm{s}_{k,i}(w^k) = \nabla_{w_k^{\sf T}} Q_k(w^k;\bm{x}_{k,i}) - \nabla_{w_k^{\sf T}} J_k(w^k)
\end{align}
and define the network vectors
\begin{align}
\bm{s}_i(w) &\triangleq \text{col}\{\bm{s}_{k,i}(w^1),...,\bm{s}_{N,i}(w^N)\} \\
\bm{Q}_i(w) &\triangleq \text{col}
\big\{\nabla_{w_1^{\sf T}} Q_1(w^1;\bm{x}_{1,i}),\dots,\notag \\
&\qquad \qquad \qquad \qquad \nabla_{w_N^{\sf T}} Q_N(w^N;\bm{x}_{N,i})\big\}
\end{align}
where we simplified the notation $\bm{s}_{k,i}(w^k)$, $\bm{s}_i(w)$ and $\bm{Q}_i(w)$ by dropping $\{\bm{x}_{k,i}\}$ from their arguments. Then, it holds that 
\begin{align}
\bm{s}_i(w) = \bm{Q}_i(w) - F(w) 
\end{align}
Note that given the action profile $w$, the randomness of $\bm{s}_{k,i}$, $\bm{s}_i$ and $\bm{Q}_i$ comes from the random data $\{\bm{x}_{k,i}\}$, and therefore we denote them in boldface. We denote by $\bm{\mathcal{F}}_{i-1}$ the collection of iterates $\{\bm{w}_{k,i-1}\}$ at all agents $k=1,...,N$ and up to time $i-1$.   
\begin{asump} ({\bf Gradient noise}) \label{assump:gradnoise}
It is assumed that the first and second-order conditional moments of the gradient noise process satisfy:
\begin{align}
	\label{gradnoise1}
	\mathbb{E}[\bm{s}_i(\bm{w}_{i-1})| \bm{\mathcal{F}}_{i-1}]&=0 \\
	\label{gradnoise2}
	\mathbb{E}\left[\|\bm{s}_i(\bm{w}_{i-1})\|^2| \bm{\mathcal{F}}_{i-1}\right]&\leq \alpha \|\bm{w}_{i-1}\|^2+\beta
\end{align}	
for some nonnegative constants $\alpha$ and $\beta$. 
\QEDB
\end{asump}
It can be verified that conditions (\ref{gradnoise1})--(\ref{gradnoise2}) are automatically satisfied for important cases of interest. For example, consider the case of quadratic losses in (\ref{IndivCost}). Some straightforward algebra shows in this case that, using stationary realizations 
$\{\bm{B}_i,\bm{b}_i\}$ for the quantities $\{B,b\}$ in (\ref{Fw}), we get the approximate block gradient vector as
\begin{align}
	\bm{Q}_i (\bm{w}_{i-1}) = \bm{B}_i \bm{w}_{i-1} + \bm{b}_i
\end{align}
so that
\begin{align}
\bm{s}_i(\bm{w}_{i-1}) &\triangleq -\widetilde{\bm{B}}_i \bm{w}_{i-1} - \widetilde{\bm{b}}_i 
\end{align}
where $\widetilde{\bm{B}}_i \triangleq B - \bm{B}_i $ and $\widetilde{\bm{b}}_i \triangleq b - \bm{b}_i$.
Note that $\mathbb{E}\widetilde{\bm{B}}_i = 0$ and $\mathbb{E}\widetilde{\bm{b}}_i= 0$ from the fact that $B=\mathbb{E} \bm{B}_i$ and $b = \mathbb{E} \bm{b}_i$. From the independence of $\bm{B}_i$, $\bm{b}_i$, and $\bm{w}_{i-1}$, Assumption~\ref{assump:gradnoise} can be seen to be satisfied since  
\begin{align}
\mathbb{E}[\bm{s}_i(\bm{w}_{i-1})| \bm{\mathcal{F}}_{i-1}] &= - \mathbb{E}[\widetilde{\bm{B}}_i] \cdot \bm{w}_{i-1} - \mathbb{E}\widetilde{\bm{b}}_i= 0  \\
\mathbb{E}\left[\|\bm{s}_i(\bm{w}_{i-1})\|^2| \bm{\mathcal{F}}_{i-1}\right] & \leq \lambda_\text{max}\!\left(\mathbb{E}[\widetilde{\bm{B}}_i^{\sf T} \widetilde{\bm{B}}_i]\right)\! \|\bm{w}_{i-1}\|^2 \!+ \mathbb{E}\|\widetilde{\bm{b}}_i\|^2 
\end{align} 
with $\alpha = \lambda_\text{max}(\mathbb{E}[\widetilde{\bm{B}}_i^{\sf T} \widetilde{\bm{B}}_i])$ and $\beta = \mathbb{E}\|\widetilde{\bm{b}}_i\|^2$.

\subsection{Stochastic Gradient Dynamics}

We consider first the stochastic-gradient implementation (\ref{StoGradient}). We can describe the evolution of the dynamics of the algorithm in terms of the aggregate quantities $\bm{w}_i \triangleq \text{col}\{\bm{w}_{1,i},...,\bm{w}_{N,i}\}$ by writing:
\begin{align}
\label{netStoGradient}
\bm{w}_i &=\bm{w}_{i-1} - U \bm{Q}_i (\bm{w}_{i-1}) - \rho U \nabla_{w^{\sf T}} p(\bm{w}_{i-1}) 
\end{align}
Subtracting $w^\star$ from both sides of (\ref{netStoGradient}), introducing the error vector $\widetilde{\bm{w}}_i \triangleq w^\star -\bm{w}_i$ and using (\ref{NashCond}) we find that
\begin{align}
\label{errStoGradient}
\widetilde{\bm{w}}_i &=\widetilde{\bm{w}}_{i-1} + U F^p(\bm{w}_{i-1}) + U \bm{s}_i(\bm{w}_{i-1})
\end{align}
The following theorem now establishes that the network error is mean-square stable for sufficiently small step-sizes $\{\mu_k\}$ and variation parameter $t$. 
\begin{thm} ({\bf Mean-square-error stability}) \label{CONVEG_GRADSTO}
	For the stochastic gradient implementation (\ref{StoGradient}), if the step-sizes $\{\mu_k\}$ satisfy
	\begin{align}
	\label{stepsize_Sto}
	0 < \mu_\text{max} < \frac{2\nu'}{(\delta + \rho\delta_p)^2+ 2\alpha}, \quad t < \frac{\nu}{\delta+\rho \delta_p} 
	\end{align} 
	then it holds that
	\begin{align}
	\lim\limits_{i \rightarrow \infty} \sup \mathbb{E}\|\widetilde{\bm{w}}_i\|^2 &= O(\mu_\text{max})
	\end{align}
\end{thm}
\begin{IEEEproof}
	See Appendix~\ref{proof:conveg_gradsto}.
\end{IEEEproof}

\subsection{Diffusion ATP and PTA Strategies}
\begin{figure}[t!]	
	\begin{minipage}[b]{1.0\linewidth}
		\centering
		\centerline{\includegraphics[width=3.1in]{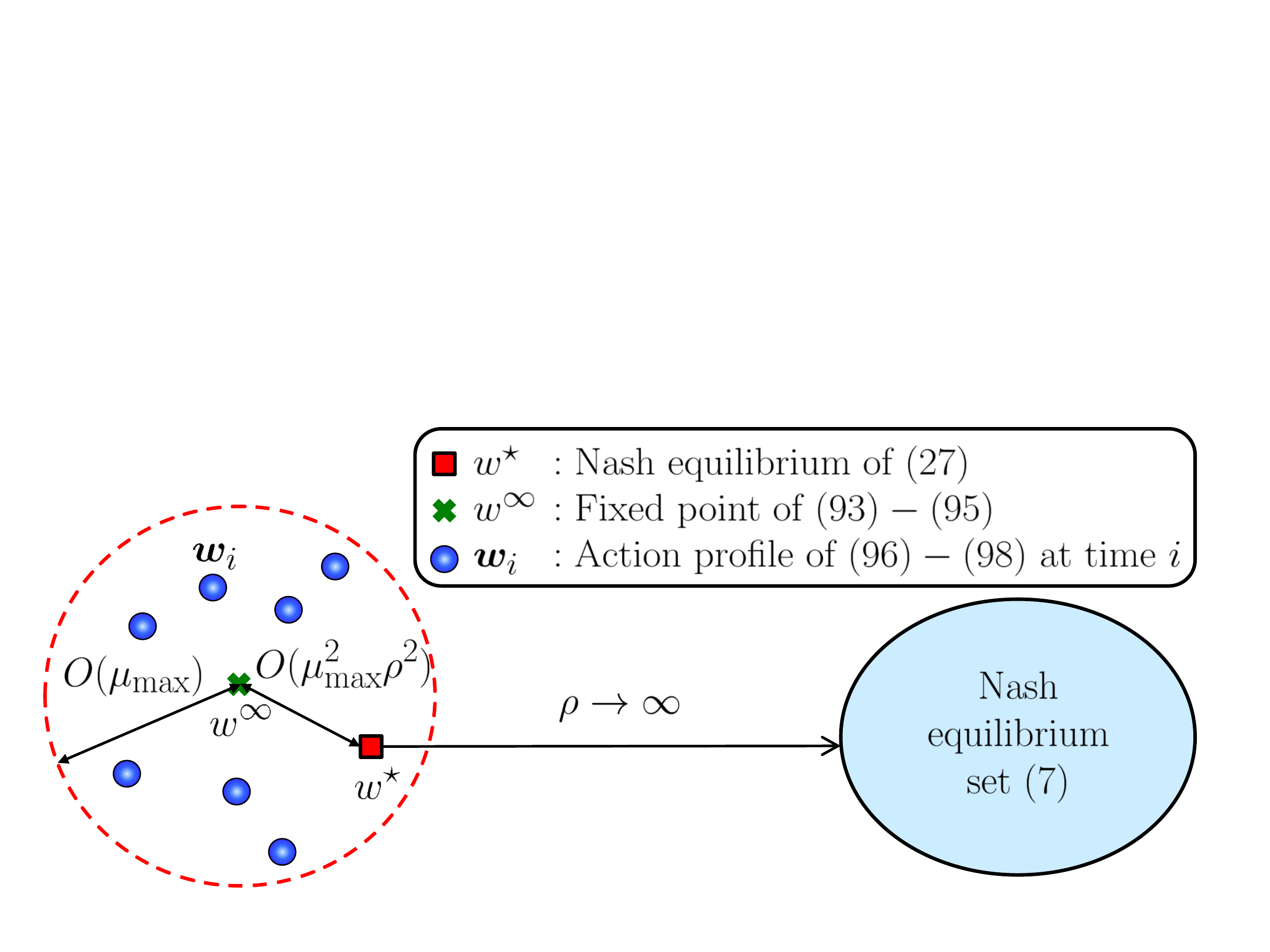}}
	\end{minipage}
	\caption{\small Illustration of the relations between $\bm{w}_i$, $w^\star$, and $w^\infty$ in steady-state for sufficiently small step-sizes. The notation $O(\mu_\text{max})$ and $O(\mu_\text{max}^2\rho^2)$ in the drawing represent the squared distances $\mathbb{E}\|w^\infty - \bm{w}_i\|^2$ and $\|w^\star-w^\infty\|$, respectively.}
	\label{wstarwinf}
\end{figure}

Let us consider next the deterministic ATP and PTA strategies
(\ref{Gradient31})--(\ref{Gradient32}) and (\ref{Gradient41})--(\ref{Gradient42}), respectively, without gradient noise. Later, we re-incorporate the gradient noise and adjust the conclusions. Thus, note that in the noiseless case we can aggregate the recursions across all agents into the following unified description:
\begin{align}
\label{netGradient20}
\phi_i&=w_{i-1} - c_1 \rho U \nabla_{w^{\sf T}} p(w_{i-1}) \\
\label{netGradient21}
\psi_i &= \phi_i - U F(\phi_i) \\
\label{netGradient22}
w_i&= \psi_i - c_2 \rho U \nabla_{w^{\sf T}} p(\psi_i)
\end{align}
for some constants $(c_1,c_2)$. By setting $(c_1,c_2)=(0,1)$ we recover the ATP recursions (\ref{Gradient31})--(\ref{Gradient32}) while for $(c_1,c_2)=(1,0)$ we obtain the PTA recursions from (\ref{Gradient41})--(\ref{Gradient42}). We thus note that the constants $(c_1,c_2)$ satisfy the properties:
\begin{align}
\label{ccc}
c_1^2=c_1, \quad c_2^2=c_2, \quad c_1 \cdot c_2 = 0, \quad c_1+c_2=1
\end{align}
The following result establishes that recursions (\ref{netGradient20})--(\ref{netGradient22}) converge to a unique fixed point.
\begin{thm} ({\bf Unique fixed point}) \label{CONVEG_GRAD2}
	The mapping from $w_{i-1}$ to $w_i$ in (\ref{netGradient20})--(\ref{netGradient22}) converges to a unique fixed point, denoted by $\psi^\infty$, for small step-sizes and for sufficiently large penalty parameters that satisfy:
	\begin{align}
	\label{stepsize_DetPen}
	0<\mu_\text{max} < \mu_o, \quad t < \frac{\nu}{\delta+\rho \delta_p}, \quad \rho > \frac{\delta}{\delta_p}
	\end{align} 
	where 
	\begin{align}
	\mu_o \triangleq \min\left\{\frac{2\nu'}{\delta^2+\rho^2 \delta_p^2-4t\nu''\rho\delta_p} ,\frac{\nu' +  \frac{t(\rho^2\delta_p^2-\delta^2)}{\rho\delta_p}}{\delta^2}\right\}
	\end{align}
\end{thm}
\begin{IEEEproof}
	See Appendix~\ref{proof:conveg_grad2}.
\end{IEEEproof}
We note that if the step-sizes are uniform, i.e., $\mu_k=\mu$ and $t=0$, the step-size condition in (\ref{stepsize_DetPen}) simplifies to
\begin{align}
0<\mu < \frac{2\nu}{\delta^2+\rho^2 \delta_p^2}
\end{align}
since
\begin{align}
\rho > \frac{\delta}{\delta_p} ~\Longleftrightarrow~ \rho^2\delta_p^2 > \delta^2 ~\Longleftrightarrow~ \frac{\nu}{\delta^2} > \frac{2\nu}{\delta^2 + \rho^2 \delta_p^2}  
\end{align}
From Theorem~\ref{CONVEG_GRAD2} we know that there exists a unique fixed point for recursion (\ref{netGradient20})--(\ref{netGradient22}), which means that we can write
\begin{align}
\label{fixGradient0}
\phi^\infty&=w^\infty - c_1 \rho U \nabla_{w^{\sf T}} p(w^\infty) \\
\label{fixGradient1}
\psi^\infty &= \phi^\infty - U F(\phi^\infty) \\
\label{fixGradient2}
w^\infty&= \psi^\infty - c_2 \rho U \nabla_{w^{\sf T}} p(\psi^\infty)
\end{align}
where we are denoting the network fixed-point vectors by $w^\infty$, $\psi^\infty$ and $\phi^\infty$.
Similarly, we can express the diffusion (stochastic) versions of the ATP and PTA strategies in (\ref{StoGradient31})--(\ref{StoGradient32}) and (\ref{StoGradient41})--(\ref{StoGradient42}) in the form:
\begin{align}
\label{netStoGradient20}
\bm{\phi}_i&=\bm{w}_{i-1} - c_1 \rho U \nabla_{w^{\sf T}} p(\bm{w}_{i-1}) \\
\label{netStoGradient21}
\bm{\psi}_i &= \bm{\phi}_i - U \bm{Q}_i(\bm{\phi}_i) \\
\label{netStoGradient22}
\bm{w}_i&=\bm{\psi}_i - c_2 \rho U \nabla_{w^{\sf T}} p(\bm{\psi}_i)
\end{align}
Let $\widetilde{\bm{w}}^\infty_i \triangleq w^\infty - \bm{w}_i$ denote the fixed-point error resulting from (\ref{netStoGradient20})--(\ref{netStoGradient22}).The following theorem shows that the variance of this error is bounded.
\begin{thm} ({\bf Bounded MSE}) \label{CONVEG_GRADSTO2}
	For the stochastic recursion  (\ref{netStoGradient20})--(\ref{netStoGradient22}), if the step-sizes $\{\mu_k\}$ and the penalty parameter $\rho$ satisfy
	\begin{align}
	\label{stepsize_StoPen}
	0 < \mu_\text{max} < \mu'_o, \quad t < \frac{\nu}{\delta+\rho \delta_p}, \quad \rho > \frac{\sqrt{\delta^2+2\alpha}}{\delta_p}
	\end{align}
	where 
	\begin{align}
	\mu'_o \triangleq \min\!\Bigg\{\!\frac{2\nu'}{\delta^2+2\alpha+\rho^2 \delta_p^2-4t\nu''\rho\delta_p}, \frac{\nu' +  \frac{t(\rho^2\delta_p^2-(\delta^2+2\alpha))}{\rho\delta_p}}{\delta^2+2\alpha} \!\Bigg\}
	\end{align}
	then it holds that for sufficiently small step-sizes
	\begin{align}
	\label{boundMSE}
	\lim\limits_{i \rightarrow \infty} \sup \mathbb{E}\|\widetilde{\bm{w}}_i^\infty\|^2  = O(\mu_\text{max})
	\end{align}	
\end{thm}
\begin{IEEEproof}
	See Appendix~\ref{proof:conveg_gradsto2}.	
\end{IEEEproof}
It is easy to verify that if the step-sizes are uniform, the step-size condition in (\ref{stepsize_StoPen}) becomes
\begin{align}
\label{unistep_StoPen}
0<\mu < \frac{2\nu}{\delta^2+2\alpha+\rho^2 \delta_p^2}
\end{align}
We note from $\alpha \geq 0$ that $\mu'_o \leq \mu_o$, which means that 
condition (\ref{stepsize_StoPen}) for the stochastic recursion implies condition (\ref{stepsize_DetPen}) for the deterministic recursion. Therefore, any $\mu_\text{max}$ satisfying (\ref{stepsize_StoPen}) ensures the existence of the fixed point $w^\infty$. 
However, the fixed point $w^\infty$ is generally different from the desired Nash equilibrium $w^\star$. In the following theorem, we examine the bias $\widetilde{w} \triangleq w^\star - w^\infty$. We show that for small $\mu_\text{max}$, the norm of the bias is asymptotically upper bounded by $O(\mu_\text{max})$. 
\begin{thm} ({\bf Small bias}) \label{SMALLBIAS}
	For sufficiently small step-sizes $\{\mu_k\}$ satisfying the following conditions:
	\begin{align}
	\label{stepsize_bias}
	0<\mu_\text{max} < \mu_o, \quad t < \frac{\nu}{\delta+\rho \delta_p}, \quad \rho > \frac{\delta}{\delta_p}
	\end{align}
	it holds that 
	\begin{align}
	\label{linearmurho}
	\lim\limits_{\mu_\text{max} \rightarrow 0} \sup \frac{\|w^\star-w^\infty\|}{\mu_\text{max}} \leq c \rho
	\end{align}
	where $c$ is a constant independent of $\mu_\text{max}$. Therefore, for sufficiently small $\mu_\text{max}$ we can write
	\begin{align}
	\lim\limits_{i \rightarrow \infty} &\sup \mathbb{E}\|w^\star-\bm{w}_i\|^2 \notag \\
	&\leq 2 \lim\limits_{i \rightarrow \infty} \sup  \mathbb{E}\|w^\infty-\bm{w}_i\|^2 + 2 \|w^\star-w^\infty\|^2 \notag \\
	& = O(\mu_\text{max}) + O(\mu_\text{max}^2 \rho^2)
	\end{align}
	
\end{thm}
\begin{IEEEproof}
	See Appendix~\ref{proof:smallbias}.
\end{IEEEproof}
In Figure~\ref{wstarwinf}, we illustrate the relation between $\bm{w}_i$, $w^\star$, and $w^\infty$ in steady-state for sufficiently small step-sizes. We note that $\bm{w}_i$, $w^\star$, and $w^\infty$ asymptotically approach to the Nash equilibrium set of the original GNEP (\ref{opt_constraint}) as $\rho \rightarrow \infty$ and $\mu_\text{max} \rightarrow 0$.
We note that condition (\ref{stepsize_StoPen}) implies conditions (\ref{stepsize_DetPen}) and (\ref{stepsize_bias}). That is, as long as the step-sizes $\{\mu_k\}$ and the penalty parameter $\rho$ satisfy (\ref{stepsize_StoPen}), the diffusion ATP and PTA learning strategies have fixed points, bounded MSE, and small bias.
Furthermore, comparing (\ref{unistep_StoPen}) with (\ref{stepsize_Sto}) we observe that by using uniform step-sizes, the diffusion ATP and PTA learning strategies are more stable than the stochastic gradient dynamic strategy (\ref{StoGradient}) since they are allowed to use a larger step-size, which would assist with faster convergence performance.
We will observe this in the simulations later. 
For the special case in Example 2, this conclusion conforms with the results in~\cite{Tu12C} that the diffusion strategies are more stable than the consensus strategy. 

\section{Case Study and Simulations}

\subsection{Stochastic Network Cournot Competition}
In this section, we consider the stochastic network Cournot competition problem~\cite{Metzler03,Contreras04,Hobbs07,Kannan12,Bimpikis14} with shared constraints. We assume that the environment is stochastically dynamic in the following manner.
Suppose that we have a network with $N$ factories, regarded as the agents discussed in this work, and $L$ markets connected to the factories. Each factory $k$ needs to determine a continuous-valued and nonnegative quantity of products to be produced and delivered to each connected market, which is defined as the action of factory $k$ denoted by $w_k = [w_k(1),...,w_k(M_k)]^{\sf T}$ where we assumed $M_k$ markets are connected to factory $k$. For each factory $k$, there exists a random quadratic production cost function to generate $\sum_{n=1}^{M_k}$ $ w_k(n)$ amount of products, i.e., the production cost function for each factory is given by
\begin{align}
	\bm{C}_{k}(w_{k}) = (x_k+ \bm{v}_{x,k}) \left(\sum_{m=1}^{M_k} w_k(m)\right)^2 
\end{align}
for some parameter $x_k>0$ and random disturbance $\bm{v}_{x,k}$ with zero mean. Furthermore, the price of products sold in each market $\ell$ is assumed to follow a linear function: 
\begin{align}
	\bm{P}_\ell(r(\ell)) = q_\ell - (y_\ell + \bm{v}_{y,\ell}) r(\ell)
\end{align}
where $q_\ell>0$ and $y_\ell>0$ are the pricing parameters, the random disturbance $\bm{v}_{y,\ell}$ is zero-mean, and $r(\ell)$ is the total amount of products delivered to market $\ell$ by all connected factories, i.e.,
\begin{align}
r(\ell) = \sum_{k=1, w_k(u) \sqsubset \ell}^N w_k(u)
\end{align}
where we write $w_k(u) \sqsubset \ell$ to represent that $w_k(u)$ is the quantity that factory $k$ delivers to market $\ell$. Note that in order to be consistent with the notation in (\ref{opt_constraint}), the index $u$ in $w_k(u)$ can be different from the index $\ell$ denoted for markets.
Consequently, each factory $k$ has an individual cost function as follows:
\begin{align}
\label{cournotcost}
J_k(w^k) 
&= \mathbb{E}\left(\bm{C}_k(w_{k}) - \sum_{\ell=1, u \sqsubset \ell}^L w_k(u) \cdot \bm{P}_\ell(r(\ell)) \right) \notag \\
&= x_k \Big(\!\sum_{m=1}^{M_k} w_k(m)\Big)^2 - \!\! \sum_{\ell=1, w_k(u) \sqsubset \ell}^L \!\!w_k(u) (q_\ell - y_\ell \cdot r(\ell))
\end{align}
Note that the loss functions in the individual cost functions can be rewritten in the quadratic form (\ref{IndivCost}). Now, let us show that $\{J_k(w^k)\}$ in the network Cournot competition are strongly monotone. For each $u \sqsubset \ell$ we have the components in $\nabla_{w_k^{\sf T}} J_k(w^k)$ as
\begin{align}
&\frac{\partial J_k(w^k)}{\partial w_k(u)} = 2 x_k \sum_{m=1}^{M_k} w_k(m) - q_\ell + y_\ell \left[ w_k(u) + r(\ell) \right]
\end{align}
If we collect these components into the long block vector $F(w)$, we get the form in (\ref{Fw}) where the $(m,n)-$th entry in each block of matrix $B$ is given by
\begin{align}
B^k_{k k}(m,n) &= \begin{cases}
x_k+y_\ell, & \text{if~} m=n \text{~s.t.~} w_k(m) \sqsubset \ell \\
x_k, & \text{if~} m \neq n 
\end{cases} \\
B^k_{k q}(m,n) &= \begin{cases}
y_\ell, & \text{if~} w_k(m) \sqsubset \ell \text{~and~} w_q(n) \sqsubset \ell \\
0, & \text{otherwise} 
\end{cases}, ~k \neq q 
\end{align}
It is easy to check that matrix $B$ can be expressed as
\begin{align}
B=X X^{\sf T} + Y_1 Y_1^{\sf T} + Y_2 Y_2^{\sf T}
\end{align}
where $X$ is an $M \times M$ diagonal matrix with diagonal entries $\{\sqrt{y_{\ell_{ku}}}\}$ for $w_k(u) \sqsubset \ell_{ku}$, $Y_1$ is a $M \times N$ block diagonal matrix with $N \times N$ blocks in which the $(k,k)-$th diagonal block is $Y_{1,kk} = \left[\sqrt{2x_{k}}, ..., \sqrt{2x_{k}}\right]^{\sf T}$ of size $M_k \times 1$, and $Y_2$ is a $M \times L$ block matrix with $N \times L$ blocks in which the $(k,\ell)-$th block is a vector of size $M_k \times 1$ and defined as 
\begin{align}
Y_{2,k \ell}(m)&= \begin{cases}
\sqrt{y_\ell}, & \text{if~}w_k(m) \sqsubset \ell \\
0, & \text{otherwise}
\end{cases}
\end{align}
Therefore, we find that $B$ has the following property for any $M \times 1$ vector $a$:
\begin{align}
a^{\sf T} B a = a^{\sf T} X X^{\sf T} a + \|Y_1^{\sf T} a\|^2 + \|Y_2^{\sf T} a\|^2 \geq x_\text{min} \cdot \|a\|^2
\end{align}
where $x_\text{min} \triangleq \min_{1\leq \ell \leq L} x_\ell$. Consequently, the network Cournot competition with individual cost functions in (\ref{cournotcost}) satisfies the strongly monotone property (\ref{strongmono}).

\begin{figure}[t!]
	\centering
	\includegraphics[width=2.9in]{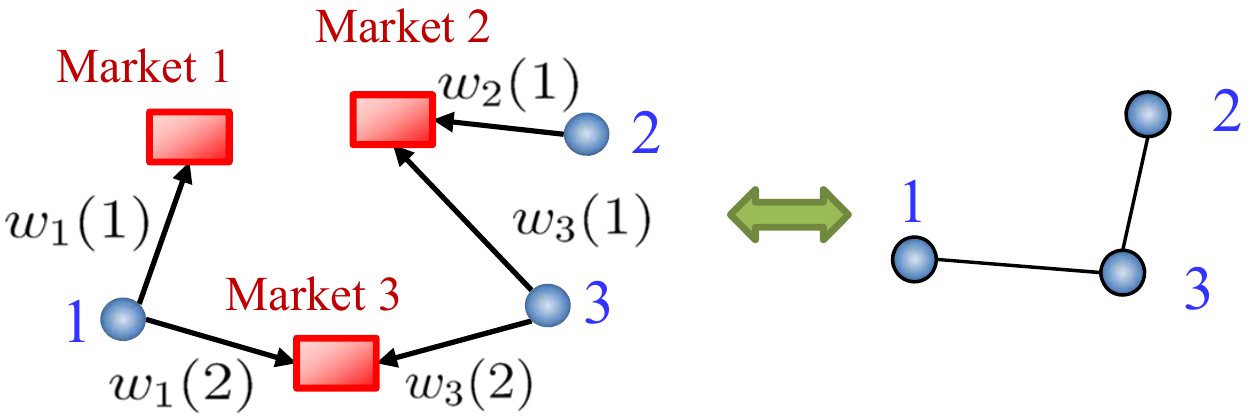}
	\caption{\small An example to illustrate the network Cournot competition and the equilivant network topology.}
	\label{cournot_eg}
\end{figure}

\begin{examp} ({\bf Cournot network with $3$ agents}) An illustrative example with $N=3$ factories and $L=3$ markets is provided in Fig.~\ref{cournot_eg}.
	Following the notations for the quantities at each link, we have the individual cost functions for the factories as
	\begin{align*}
	J_1(w^1) &= x_1 \left[ w_1(1) + w_1(2)\right]^2 - w_1(1) \cdot \left(q_1 - y_1 \cdot w_1(1)\right) \notag \\
	&~~~ - w_1(2) \cdot \left(q_2 - y_2 \cdot [w_1(2) + w_3(2)]\right)  \\
	J_2(w^2) &= x_2 [w_2(1)]^2 - w_2(1) \cdot \left(q_2 - y_2 \cdot [w_1(2) + w_3(2)]\right)  \\
	J_3(w^3) &= x_3 \left[ w_3(1) + w_3(2)\right]^2 \notag \\
	&~~~ - w_3(1) \cdot \left(q_2 - y_2 \cdot [w_1(2) + w_3(2)]\right) \notag \\
	&~~~ - w_3(2) \cdot \left(q_3 - y_3 \cdot [w_2(1) + w_3(1)]\right) 
	\end{align*}
	Therefore, we get
	\begin{align*}
	\frac{\partial J_1(w^1)}{\partial w_1(1)} &= 2 x_1 \left[ w_1(1) + w_1(2)\right] - q_1 + 2 y_1 w_1(1) \\
	\frac{\partial J_1(w^1)}{\partial w_1(2)} &= 2 x_1 \left[ w_1(1) + w_1(2)\right] - q_3 + 2 y_3 w_1(2) + y_3 w_3(2) \\
	\frac{\partial J_2(w^2)}{\partial w_2(1)} &= 2 x_2 w_2(1) - q_2 + 2 y_2 w_2(1) + y_2 w_3(1) \\
	\frac{\partial J_3(w^3)}{\partial w_3(1)} &= 2 x_3 \left[ w_3(1) + w_3(2)\right] - q_2 + 2 y_2 w_3(1) + y_2 w_2(1) \\
	\frac{\partial J_3(w^3)}{\partial w_3(2)} &= 2 x_3 \left[ w_3(1) + w_3(2)\right] - q_3 + 2 y_3 w_3(2) + y_3 w_1(2) 
	\end{align*}
	It can be then verified that matrix $B$ is given by
	\begin{align}
	B&=\begin{bmatrix}
	2 x_1 + 2 y_1 &\hspace{-2mm} 2 x_1 &\hspace{-2mm} 0 &\hspace{-2mm} 0 &\hspace{-2mm} 0 \\
	2 x_1 &\hspace{-2mm} 2 x_1 + 2 y_3 &\hspace{-2mm} 0 &\hspace{-2mm} 0 &\hspace{-2mm} y_3 \\
	0 &\hspace{-2mm} 0 &\hspace{-2mm} 2 x_2 + 2 y_2 &\hspace{-2mm} y_2 &\hspace{-2mm} 0 \\
	0 &\hspace{-2mm} 0 &\hspace{-2mm} y_2 &\hspace{-2mm} 2 x_3 + 2 y_2 &\hspace{-2mm} 2 x_3 \\
	0 &\hspace{-2mm} y_3 &\hspace{-2mm} 0 &\hspace{-2mm} 2 x_3 &\hspace{-2mm} 2 x_3 + 2 y_3
	\end{bmatrix} \notag \\
	&= X X^{\sf T} + Y_1 Y_1^{\sf T} + Y_2 Y_2^{\sf T}
	\end{align}
	where
	\begin{align}
	X &= \text{diag}\{\sqrt{y_1}, \sqrt{y_3}, \sqrt{y_2}, \sqrt{y_2}, \sqrt{y_3}\} \\
	Y_1 &= \begin{bmatrix}
	\sqrt{2x_1} & \sqrt{2x_1} & 0 & 0 & 0 \\
	0 & 0 & \sqrt{2x_2} & 0 & 0\\
	0 & 0 & 0 & \sqrt{2x_3} & \sqrt{2x_3} 
	\end{bmatrix}^T \\
	Y_2 &= \begin{bmatrix}
	\sqrt{y_1} & 0 & 0 & 0 & 0 \\
	0 & 0 & \sqrt{y_2} & \sqrt{y_2} & 0\\
	0 & \sqrt{y_3} & 0 & 0 & \sqrt{y_3}
	\end{bmatrix}^T
	\end{align}
	\QEDB
\end{examp}

\subsection{Numerical Results}
\label{cournotnet}

\begin{figure}[t!]
	\centering
	\includegraphics[width=1.8in]{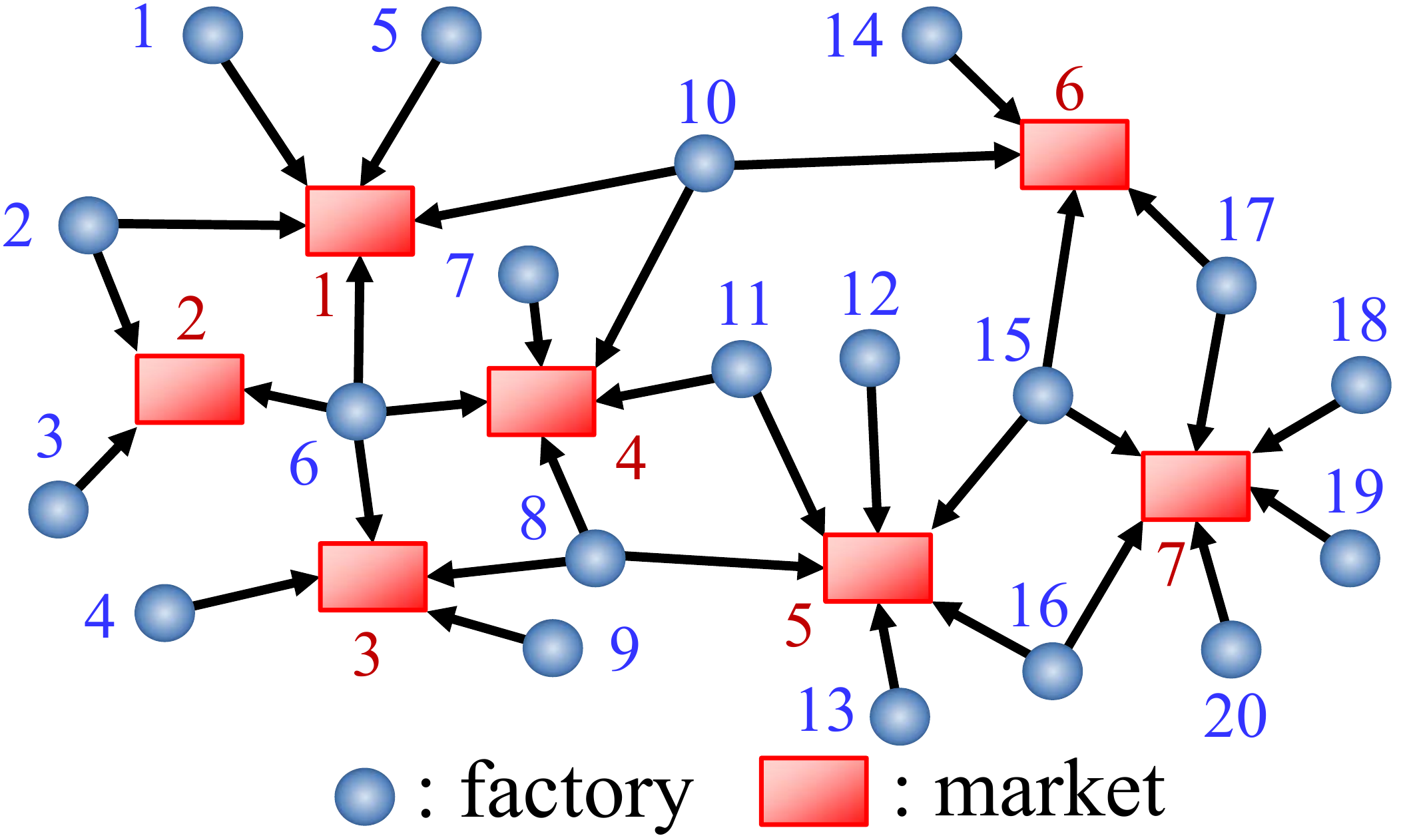}
	\caption{\small Network structure used for the simulations of the network Cournot competition.}
	\label{cournot_sim}
\end{figure}

In the simulations, we consider a network with $N=20$ factories and $L=7$ markets which are connected as shown in Fig.~\ref{cournot_sim}. For each individual cost function $J_k(w^k)$, we set ${x_k=4}$, $q_\ell=12$, and $y_\ell=4$ for all $k$ and $\ell$ in (\ref{cournotcost}). 
For the stochastic setting, the realizations of random noises $\bm{v}_{x,k}$ and $\bm{v}_{y,\ell}$ for all $k$ and $\ell$ are generated at each time instant $i$, and are assumed to be temporally and spatially independent. 
We further assume that both $\bm{v}_{x,k}$ and $\bm{v}_{y,\ell}$ are uniformly distributed between $[-4,4]$. The step-sizes are assumed to be uniform, i.e., $\mu_k=\mu$ for all $k$.

The action $w_k$ of each factory $k$ needs to be determined under the following constraints. The quantity of products delivered to each market has to be nonnegative and each market $\ell$ has an upper limit capacity  $h_\ell$ of products, i.e., for $m=1,...,M_k$ and $\ell=1,...,L$,
\begin{align}
\label{courconst}
	w_k(m) \geq 0, \quad r(\ell) = \sum_{k=1, w_k(u) \sqsubset \ell}^N w_k(u) \leq h_\ell
\end{align}
where $h_\ell$ is set to be $1$ in the experiments. Furthermore, we apply the quadratic penalty function in (\ref{penalty}) to each constraint in the algorithms. We remark that the proposed penalty methods give only asymptotically feasible solutions, which could be improved by imposing harsher penalty or considering stricter constraints than (\ref{courconst}). However, we rely on (\ref{courconst}) in the simulations to examine the numerical performance regardless of solution feasibility.

\begin{figure}[t!]
	\centering
	\includegraphics[width=2.5in]{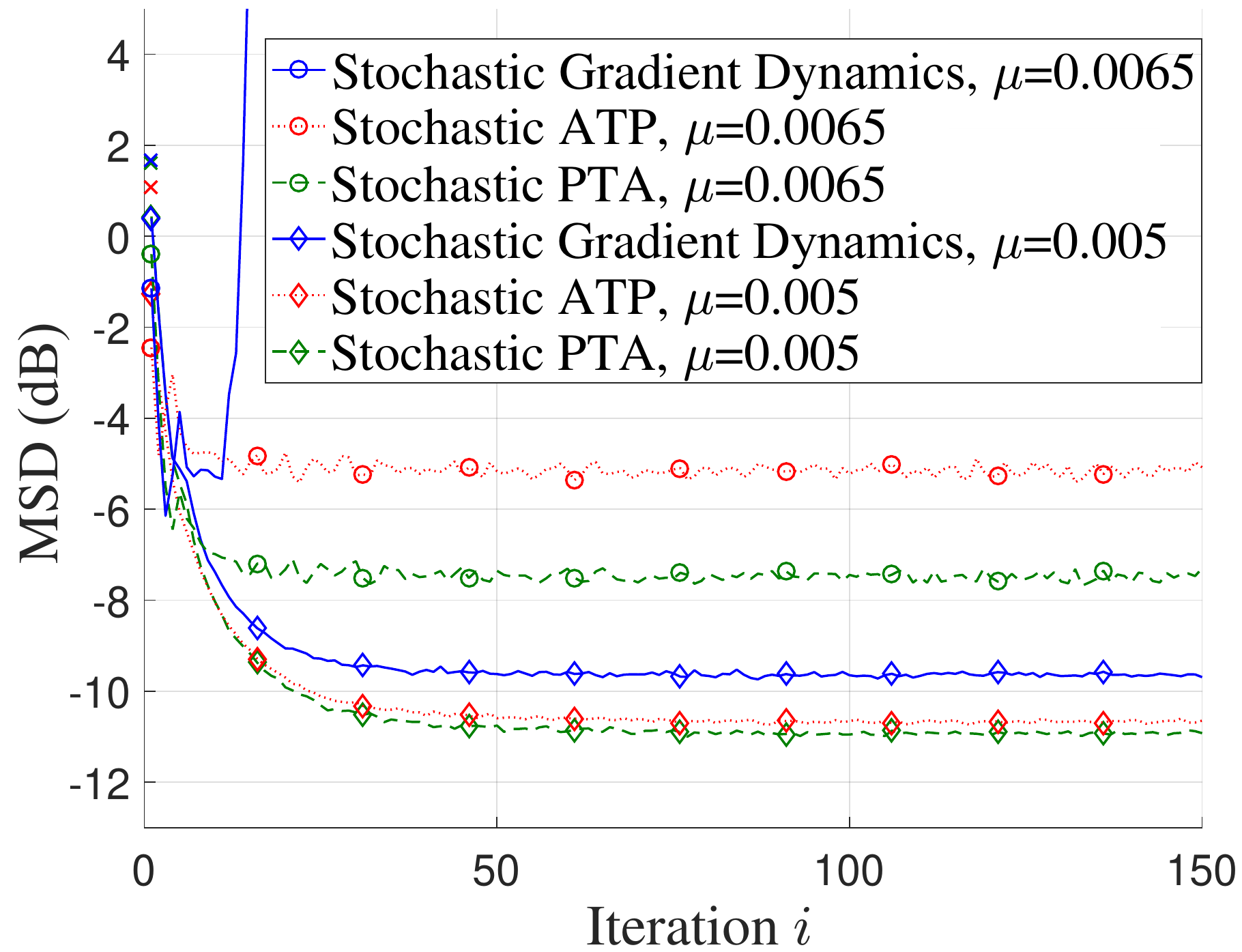}
	\caption{\small MSD learning curves for the stochastic gradient dynamic, diffusion ATP, and diffusion PTA with different step-sizes $\mu$.}
	\label{diffmu}
\end{figure}

\begin{figure}[t!]
	\centering
	\includegraphics[width=2.5in]{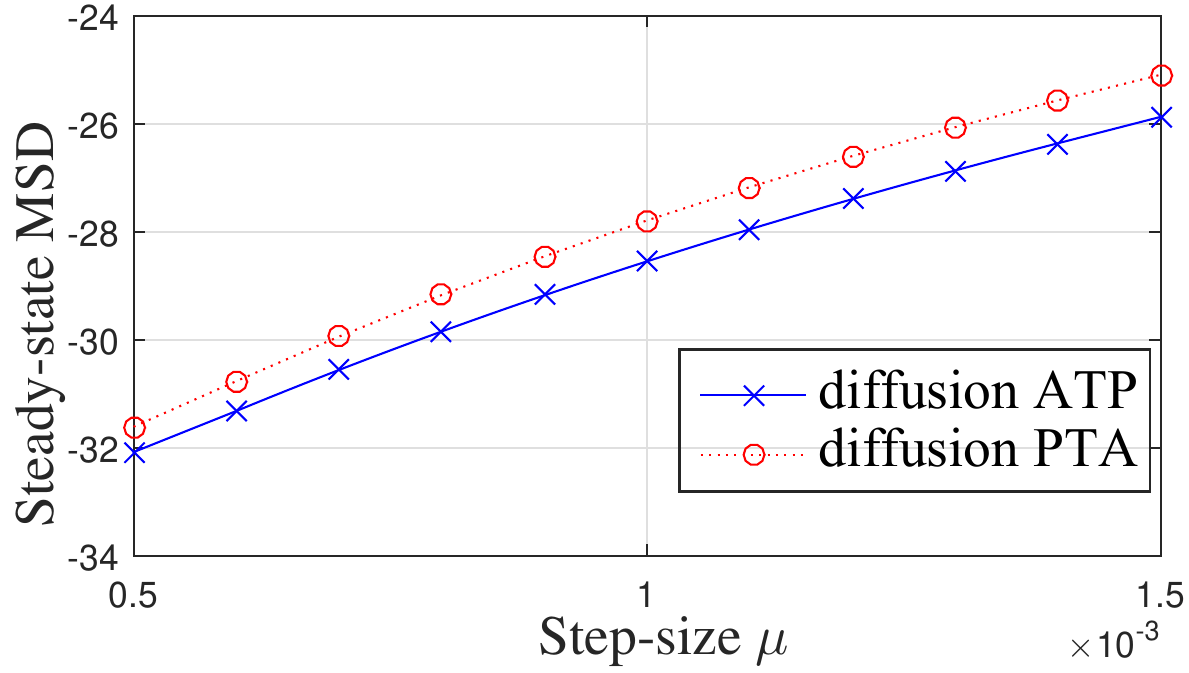}
	\caption{\small The steady-state MSD for diffusion ATP and diffusion PTA.}
	\label{MSEbound}
\end{figure}

\begin{figure}[t!]
	\centering
	\includegraphics[width=3in]{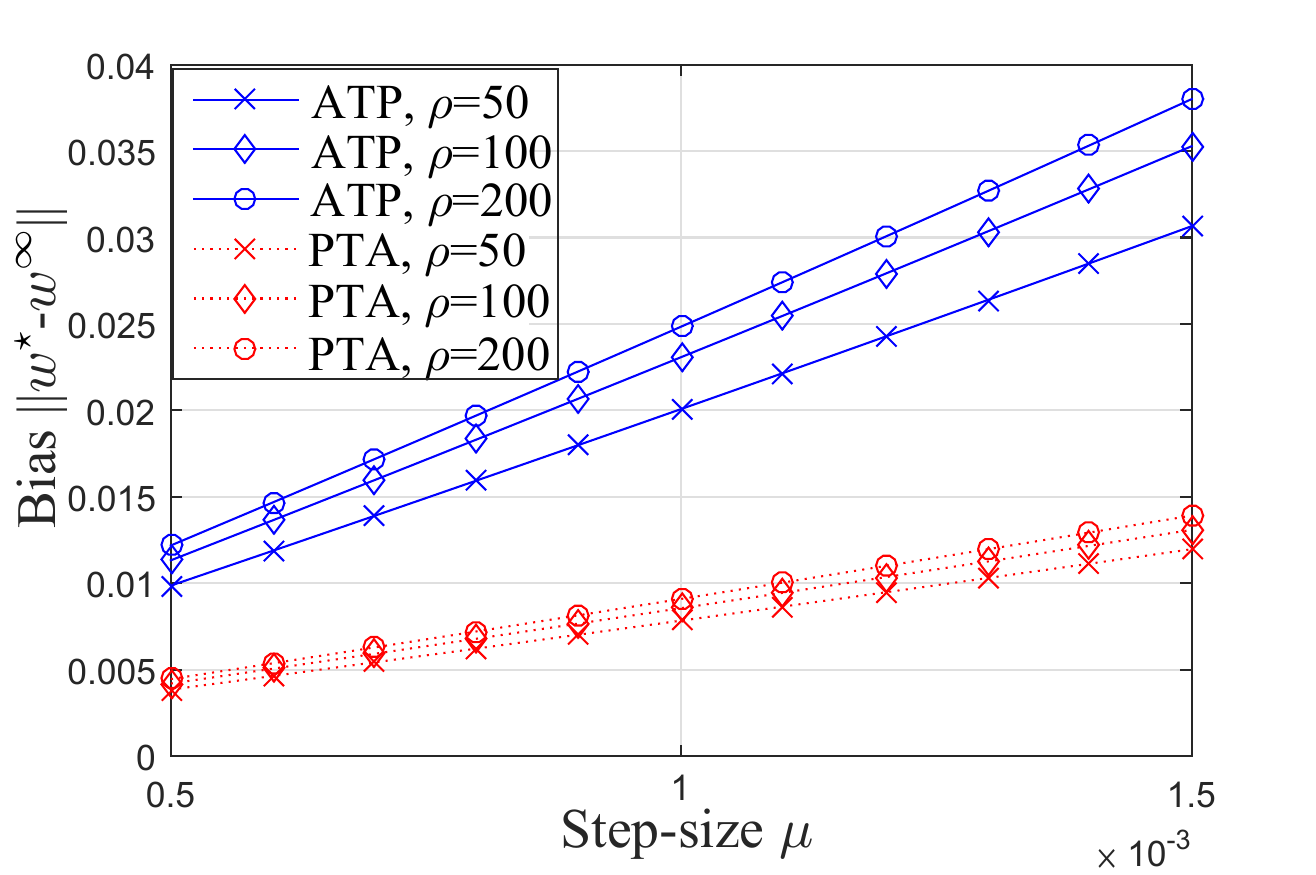}
	\caption{\small The bias distance $\|w^\star-w^\infty\|$ from the Nash equilibrium $w^\star$ to fixed points $w^\infty$ for diffusion ATP and diffusion PTA.}
	\label{bias}
\end{figure}

We first set the penalty parameter $\rho$ to $200$ and vary the step-size $\mu$ for the stochastic gradient dynamic (\ref{StoGradient}), ATP strategy (\ref{StoGradient31})-(\ref{StoGradient32}), and PTA strategy (\ref{StoGradient41})-(\ref{StoGradient42}). In Fig.~\ref{diffmu}, we study the mean-square-deviation (MSD) performance, defined as $\mathbb{E}\|w^\infty - \bm{w}_i\|^2$, for each algorithm toward its fixed point. Note that for the stochastic gradient case we have $w^\infty=w^\star$. We can see that with a smaller step-size $\mu$, the three algorithms exhibit smaller steady-state MSD values while converging slower, and their differences vanish with smaller $\mu$ as well.
It is worthwhile to note though that the diffusion ATP and PTA strategies generally outperform the stochastic gradient dynamic. Furthermore, the ATP and PTA strategies allow larger ranges of step-sizes, as we can see that for $\mu=0.0065$ these two strategies converge while the stochastic gradient dynamic does not. In Fig.~\ref{MSEbound}, we observe that for sufficiently small step-sizes, the steady-state MSD values of diffusion ATP and PTA decrease linearly with respect to $\mu$, as we expect from (\ref{boundMSE}). The bias between the fixed points $w^\infty$ and the Nash equilibrium $w^\star$ is shown in Fig.~\ref{bias}. We can see that the bias $\|w^\star-w^\infty\|$ is linear with respect to the step-size $\mu$ and the slope becomes steep when $\rho$ increases, which verifies the result in (\ref{linearmurho}). Comparing diffusion ATP and PTA strategies using sufficiently small step-sizes, we find that diffusion ATP exhibits smaller steady-state MSD values than diffusion PTA; on the other hand, diffusion PTA shows smaller bias values than diffusion ATP. This result would depend on the structure of the individual costs and the shared constraints, and the selection of the penalty functions $\theta_\text{IP}$ and $\theta_\text{EP}$. However, as the step-size decreases, the difference between diffusion ATP and PTA strategies becomes small in terms of the steady-state MSD and bias.

\begin{figure}[t!]
	\centering
	\includegraphics[width=2.7in]{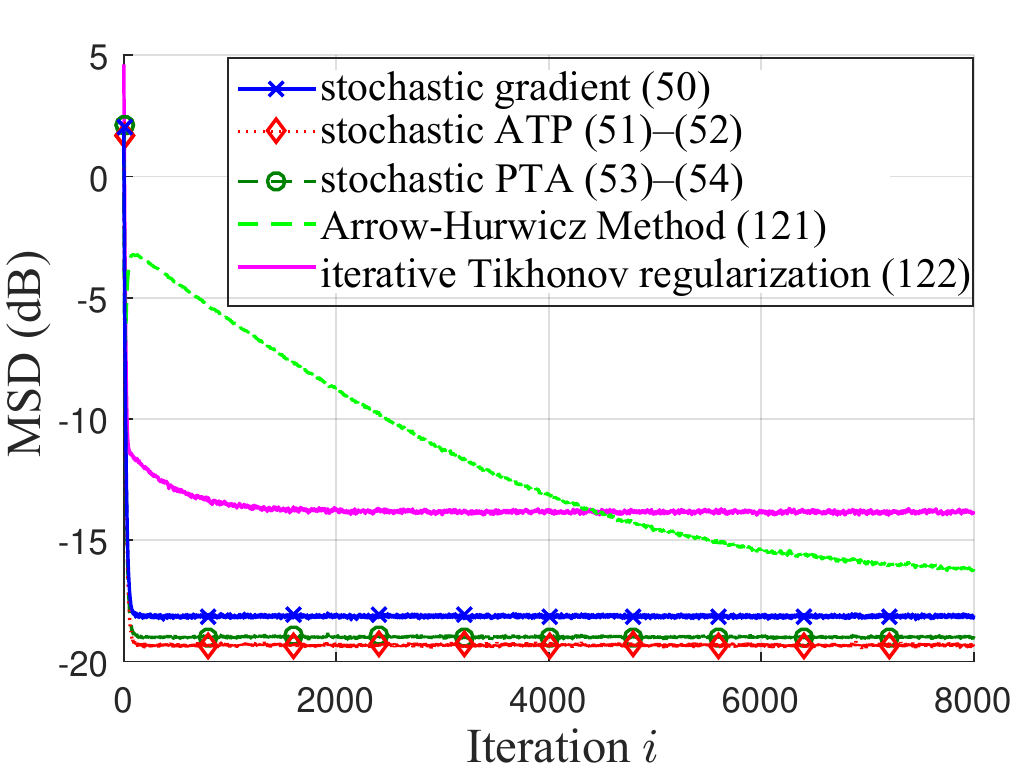}
	\caption{\small Comparisons of MSD learning curves for algorithms.}
	\label{compare}
\end{figure}

For comparisons, we simulate two related projection-based stochastic algorithms discussed in~\cite{Koshal13}, i.e., the distributed Arrow-Hurwicz method and the iterative Tikhonov regularization. Both algorithms use a constant and uniform step-size $\mu = 0.003$ in our setting. The distributed Arrow-Hurwicz method consists of the following two steps:
\begin{align}
	\begin{cases}
	\bm{w}_{k,i} = \Pi_{\mathbb{R}^+} \Big[\bm{w}_{k,i-1} - \mu \Big(\widehat{\nabla_{w_k^{\sf T}}} J_k (\bm{w}_{k,i-1}) \\
	\qquad \qquad \qquad \qquad \qquad \quad \sum_{\ell=1}^{L} \bm{\lambda}_{\ell,i-1} (\bm{r}_i(\ell)-h_\ell) \Big) \Big] \\
	\bm{\lambda}_{\ell,i} = \Pi_{\mathbb{R}^+} \left[ \bm{\lambda}_{\ell,i-1} + \mu (\bm{r}_i(\ell)-h_\ell ) \right]
	\end{cases}	
\end{align}
where $\bm{r}_i(\ell)$ denotes the random realization for $r(\ell)$ at time $i$. On the other hand, the iterative Tikhonov regularization follows these two steps:
\begin{align}
	\begin{cases}
	\bm{w}_{k,i} = \Pi_{\mathbb{R}^+} \Big[\bm{w}_{k,i-1} - \mu \Big(\epsilon_i \bm{w}_{k,i-1} + \widehat{\nabla_{w_k^{\sf T}}} J_k (\bm{w}_{k,i-1}) \\
	\qquad \qquad \qquad \qquad \qquad \qquad \sum_{\ell=1}^{L} \bm{\lambda}_{\ell,i-1} (\bm{r}_i(\ell)-h_\ell) \Big) \Big] \\
	\bm{\lambda}_{\ell,i} = \Pi_{\mathbb{R}^+} \left[ \bm{\lambda}_{\ell,i-1} + \mu (\bm{r}_i(\ell)-h_\ell )- \mu \epsilon \bm{\lambda}_{\ell,i-1} \right]
\end{cases}	
\end{align}
where $\epsilon = 0.5012$ is the regularization parameter. We note that these two algorithms rely on the additional use of $L$ Lagrange multiplier(s) to deal with the shared constraints, which require some additional ``bridge nodes'' for implementation. Furthermore, the projection step incurs additional computation complexity. These two problems do not appear in our penalty-based algorithms proposed in this work. In Fig.~\ref{compare}, we simulate the MSD learning curves for these algorithms. In order to make a fair comparison, we set the step-size $\mu=0.003$ for the penalty-based strategies. The penalty parameter $\rho$ is set to 200. We observe that the stochastic gradient dynamic, ATP, and PTA strategies converge much faster than the distributed Arrow-Hurwicz method and the iterative Tikhonov regularization. Furthermore, the distributed Arrow-Hurwicz and the iterative Tikhonov regularization methods have larger steady-state MSD values than the three penalty-based algorithms.

\section{Conclusion}

This work focuses on GNEPs with shared constraints over network topologies in stochastic environments. We develop three fully-distributed online learning strategies which asymptotically approach the set of generalized Nash equilibrium for small constant step-sizes and sufficiently large penalty parameters. An interesting future work would be to explore how the converging point of our algorithms in the set of GNE(s) relate to the variational equilibrium obtained by KKT conditions with identical Lagrange multipliers~\cite{Kulkarni12,Kulkarni122}. Another possibility for future work is to explore the use of sub-gradient methods would be useful for sub-differentiable penalty functions and/or individual cost functions~\cite{Ying16,Nedic09,Nesterov09}. Asynchronous adaptation learning~\cite{Zhao15,Nassif16} is also a useful extension so that agents do not need to execute the update of actions simultaneously.

\appendices

\section{Proof of Theorem~\ref{UNIQUENESS}} \label{proof:uniqueness}
We introduce the aggregate penalty function
\begin{align} 
p(w) \triangleq \sum_{u=1}^{U} \theta_\text{EP}(h_u(w)) + \sum_{q=1}^{L} \theta_\text{IP}(g_q(w))
\end{align} 
and note that 
\begin{align}  
\label{gradientpw}
\nabla_{w_k^{\sf T}} \;p(w) &= \sum_{u=1}^{U} \nabla_{h_u(w)} \theta_\text{EP}(h_u(w)) \cdot \nabla_{w_k^{\sf T}} h_u(w) \notag \\
&~~~ + \sum_{q=1}^{L} \nabla_{g_q(w)} \theta_\text{IP}(g_q(w)) \cdot \nabla_{w_k^{\sf T}} g_q(w) 
\end{align}  
and
\begin{align}  
\label{gradientpwk}
\nabla_{w_k^{\sf T}} \;p_k(w^k) &= \sum_{u=1}^{U_k} \nabla_{h_{k,u}(w^k)} \theta_\text{EP}(h_{k,u}(w^k)) \cdot \nabla_{w_k^{\sf T}} h_{k,u}(w^k) \notag \\
&~~~ + \sum_{q=1}^{L_k} \nabla_{g_{k,q}(w^k)} \theta_\text{IP}(g_{k,q}(w^k)) \cdot \nabla_{w_k^{\sf T}} g_{k,q}(w^k)
\end{align} 
Recall that, as defined in (\ref{distinctpenalty}), the global constraint functions $\{h_u(w)\}$ and $\{g_q(w)\}$ are distinctly collected and include all $\{h_{k,u}(w^k)\}$ and $\{g_{k,q}(w^k)\}$ in the network. Therefore, if a global constraint function $h_u(w)$ or $g_q(w)$ relates to some action $w_k$, agent $k$ is subject to the same constraint function, say, $h_{k,u'}(w^k)=h_u(w)$ or $g_{k,q'}(w^k)=g_q(w)$. That is, we can find one-to-one mapping from every nonzero $\nabla_{w_k^{\sf T}} h_u(w)$ or $\nabla_{w_k^{\sf T}} g_q(w)$ in (\ref{gradientpw}) to some $\nabla_{w_k^{\sf T}} h_{k,u'}(w^k)$ or $\nabla_{w_k^{\sf T}} g_{k,q'}(w^k)$ in (\ref{gradientpwk}), which means that we have
\begin{align} 
\nabla_{w_k^{\sf T}} p_k(w^k) = \nabla_{w_k^{\sf T}} \;p(w)
\end{align}   
so that 
\begin{align} 
\nabla_{w^{\sf T}} \; p(w) = \text{col}\{ \nabla_{w_1^{\sf T}} p_1(w^1),...,\nabla_{w_N^{\sf T}} p_N(w^N) \}
\end{align}   
From (\ref{Fw}) and (\ref{NashCond}) we can write 
\begin{align} 
\label{Fp}
F^p(w) & \triangleq \text{col}\{\nabla_{w_1^{\sf T}} J_1^p(w^1),...,\nabla_{w_N^{\sf T}} J_N^p(w^N)\} \notag \\
&=F(w) + \rho \nabla_{w^{\sf T}} p(w) 
\end{align} 
Since the sum of convex functions is also convex, we know that $p(w)$ is convex and, therefore, for any $w^a$ and $w^b$:
\begin{align} 
\label{conv_p}
(w^a - w^b)^{\sf T} [\nabla_{w^{\sf T}} p(w^a)-\nabla_{w^{\sf T}} p(w^b)] \geq 0
\end{align}   
Using (\ref{strongmono}) we get
\begin{align} 
\label{strongmono2}
&(w^a - w^b)^{\sf T} [F^p(w^a)-F^p(w^b)] \notag \\
&~=(w^a - w^b)^{\sf T} [F(w^a) - F(w^b) + \nabla_{w^{\sf T}} p(w^a)-\nabla_{w^{\sf T}} p(w^b)] \notag \\
&~\geq \nu \|w^a-w^b\|^2
\end{align} 
It follows that the penalized mapping $F^p: \mathbb{R}^M \rightarrow \mathbb{R}^M$ is strongly monotone. 
In order to examine the existence of a Nash equilibrium, we need to show that the strong monotonicity of $F^p(w)$ satisfies the coerciveness property~\cite[p. 14]{Kinderlehrer82},
i.e., for some $w^\text{ref}\in \mathbb{R}^M$, 
\begin{align} 
\label{coercive}
\lim\limits_{\|w\|\rightarrow \infty}
\frac{\left[F^p(w)-F^p(w^\text{ref})\right]^{\sf T} (w - w^\text{ref})}{\|w-w^\text{ref}\|} = \infty
\end{align}   
Using (\ref{strongmono2}) and setting $w^a=w$ and $w^b=w^\text{ref}$ we get
\begin{align} 
&\lim\limits_{\|w\|\rightarrow \infty}
\frac{\left[F^p(w)-F^p(w^\text{ref})\right]^{\sf T} (w - w^\text{ref})}{\|w-w^\text{ref}\|} \notag \\
&~~~\geq \lim\limits_{\|w\|\rightarrow \infty}  \nu\|w-w^\text{ref}\| \notag \\
&~~~\geq \lim\limits_{\|w\|\rightarrow \infty}  \nu\|w\|-\nu\|w^\text{ref}\| \notag \\
&~~~ = \infty
\end{align}   
which shows that $F^p(w)$ satisfies the coerciveness property in (\ref{coercive}) with $w^\text{ref}$. 
We then conclude the existence of solutions to problem (\ref{opt_reform}).

The uniqueness of the Nash equilibrium is also guaranteed by the strong monotonicity following~\cite[Theorem 2.3.3]{Facchinei03}. Since $J^p_k(w_k;w_{-k})$ is convex and differentiable in $w_k$, from the optimality criterion~\cite{Boyd04} we know that the Nash equilibrium satisfies
\begin{align}  
\left(w'_k-w^\star_k\right)^{\sf T} \nabla_{w_k} J^p_k(w^\star_k;w^\star_{-k}) \geq 0
\end{align}    
for all feasible $w'_k$. Summing up these conditions over all agents we get
\begin{align}  
\sum_{k=1}^{N}\left(w'_k-w^\star_k\right)^{\sf T} \nabla_{w_k} J^p_k(w^\star_k;w^\star_{-k}) = (w'-w^\star)^{\sf T} F^p(w^\star) \geq 0
\end{align}    
Let us first assume the existence of two distinct solutions, $w^\star\neq w^\dagger \in \mathbb{R}^M$. Then, for any $w'\in \mathbb{R}^M$ they will satisfy 
\begin{align}  
(w'-w^\star)^{\sf T} F^p(w^\star) \geq 0,~~ (w'-w^\dagger)^{\sf T} F^p(w^\dagger) \geq 0
\end{align}    
Setting $w'=w^\dagger$ in the first inequality and $w' = w^\star$ in the second inequality we get 
\begin{align}  
(w^\dagger-w^\star)^{\sf T} F^p(w^\star) \geq 0,~~ (w^\star-w^\dagger)^{\sf T} F^p(w^\dagger) \geq 0
\end{align}    
By adding these two inequalities, we arrive at
\begin{align}  
(w^\dagger-w^\star)^{\sf T} [F^p(w^\dagger)-F^p(w^\star)] \leq 0
\end{align}     
which contradicts the strong monotonicity of $F^p(w)$. We thus conclude that the Nash equilibrium is unique.
Now, from the optimality criterion~\cite{Boyd04} and given $w^\star_{-k}$, we note that $w^\star_k$ is optimal if, and only if,
\begin{align} 
\nabla_{w_k} J^p_k(w^\star_k;w^\star_{-k}) = 0
\end{align}    
Collecting these conditions for all agents we obtain
\begin{align} 
F^p(w^\star) = F(w^\star) + \rho \nabla_{w^{\sf T}} p(w^\star) = 0 
\end{align}

\section{Proof of Lemma~\ref{LIP_CONT}} \label{proof:Lip_cont}

Using Condition~\ref{cond:LipschitzPw}, we have
\begin{align} 
\label{Lippenalfunc}
\| &\nabla_{w^{\sf T}} p(w^\circ)- \nabla_{w^{\sf T}} p(w^\bullet)\|^2 \notag \\
&= \sum\limits_{k=1}^{N} \left\| \nabla_{w_k^{\sf T}} p_k(w_\circ^k)-\nabla_{w_k^{\sf T}} p_k(w_\bullet^k) \right\|^2 \notag \\
& \leq \sum\limits_{k=1}^{N} \gamma_k^2 \|w_\circ^k-w_\bullet^k\|^2 \notag \\
& \leq \delta_p^2 \|w^\circ-w^\bullet\|^2
\end{align}    
where we used the fact that $\|w_\circ^k-w_\bullet^k\|^2 \leq \|w^\circ-w^\bullet\|^2$. Then, it follows that
\begin{align} 
\label{Lippenal}
\|F^p&(w^\circ) - F^p(w^\bullet)\| \notag \\
& \leq \|F(w^\circ)-F(w^\bullet)\|+ \rho \| \nabla_{w^{\sf T}} p(w^\circ)- \nabla_{w^{\sf T}} p(w^\bullet)\| \notag \\
& \leq (\delta+\rho \delta_p) \|w^\circ-w^\bullet\|
\end{align}    
as claimed.

\section{Proof of Lemma~\ref{heterstrong}} \label{proof:heterstrong}

We first note that
\begin{align}
&(w^\circ - w^\bullet)^{\sf T} U [F^p(w^\circ) - F^p(w^\bullet)] \notag \\
&= \sum_{k=1}^{N} \mu_k (w^\circ_k - w^\bullet_k)^{\sf T} \left[\nabla_{w_k^{\sf T}} J_k^p (w_\circ^k)-\nabla_{w_k^{\sf T}} J_k^p(w_\bullet^k) \right] \notag \\
&= \mu_\text{max} \sum_{k=1}^{N} (w^\circ_k - w^\bullet_k)^{\sf T} \left[\nabla_{w_k^{\sf T}} J_k^p (w_\circ^k)-\nabla_{w_k^{\sf T}} J_k^p(w_\bullet^k) \right] \notag \\
&~~- \sum_{k=1}^{N} (\mu_\text{max}-\mu_k) (w^\circ_k - w^\bullet_k)^{\sf T} \left[\nabla_{w_k^{\sf T}} J_k^p (w_\circ^k)-\nabla_{w_k^{\sf T}} J_k^p(w_\bullet^k) \right] 
\end{align}    
Using the Cauchy-Schwartz inequality we get
\begin{align}
&\sum_{k=1}^{N} (\mu_\text{max}-\mu_k) (w^\circ_k - w^\bullet_k)^{\sf T} \left[\nabla_{w_k^{\sf T}} J_k^p (w_\circ^k)-\nabla_{w_k^{\sf T}} J_k^p(w_\bullet^k) \right] \notag \\
& \leq (\mu_\text{max}-\mu_\text{min}) \sum_{k=1}^{N}
(w^\circ_k - w^\bullet_k)^{\sf T} \left[\nabla_{w_k^{\sf T}} J_k^p (w_\circ^k)-\nabla_{w_k^{\sf T}} J_k^p(w_\bullet^k) \right] \notag \\
\label{CauH}
& \leq t\mu_\text{max} \sum_{k=1}^{N}
\|w^\circ_k - w^\bullet_k\| \cdot \left\|\nabla_{w_k^{\sf T}} J_k^p (w_\circ^k)-\nabla_{w_k^{\sf T}} J_k^p(w_\bullet^k) \right\| \notag \\
& \stackrel{(a)}{\leq} t\mu_\text{max} \left(\sum_{k=1}^{N}
\|w^\circ_k - w^\bullet_k\|^2\right)^\frac{1}{2} \notag \\
&\qquad~~~~ \times \left(\sum_{k=1}^{N} \left\|\nabla_{w_k^{\sf T}} J_k^p (w_\circ^k)-\nabla_{w_k^{\sf T}} J_k^p(w_\bullet^k) \right\|^2 \right)^\frac{1}{2} \notag \\
& = t\mu_\text{max} 
\|w^\circ - w^\bullet\| \cdot \left\|F^p (w^\circ)-F^p(w^\bullet) \right\|
\end{align}    
where $(a)$ is obtained from H{\"o}lder's inequality~\cite{Horn90}.
By (\ref{CauH}) we have
\begin{align}
(w^\circ - &w^\bullet)^{\sf T} U[F^p(w^\circ)-F^p(w^\bullet)] \notag \\
& \geq \mu_\text{max} (w^\circ - w^\bullet)^{\sf T}  [F^p(w^\circ) - F^p(w^\bullet)] \notag \\
& ~~~- t\mu_\text{max} 
\|w^\circ - w^\bullet\| \cdot \left\|F^p (w^\circ)-F^p(w^\bullet) \right\| \notag \\
& \geq \mu_\text{max} \nu \|w^\circ - w^\bullet\|^2 - t\mu_\text{max} (\delta+\rho \delta_p) \|w^\circ - w^\bullet\|^2 \notag \\
& = \mu_\text{max} [\nu - t (\delta+\rho \delta_p)]\cdot \|w^\circ - w^\bullet\|^2
\end{align}    
where we used the strong monotonicity property (\ref{strongmono2}) and the Lipschitz continuous property (\ref{Lippenal}). Similarly, we can express
\begin{align}
(w^\circ - &w^\bullet)^{\sf T} U [F(w^\circ) - F(w^\bullet)] \notag \\
& \geq \mu_\text{max} (w^\circ - w^\bullet)^{\sf T}  [F(w^\circ) - F(w^\bullet)] \notag \\
& ~~~- t\mu_\text{max} 
\|w^\circ - w^\bullet\| \cdot \left\|F(w^\circ) - F(w^\bullet) \right\| \notag \\
& \geq \mu_\text{max}(\nu- t\delta) \|w^\circ - w^\bullet\|^2 
\end{align}    
and
\begin{align}
(w^\circ - &w^\bullet)^{\sf T} U [\nabla_{w^{\sf T}} p(w^\circ) - \nabla_{w^{\sf T}} p(w^\bullet)] \notag \\
& \geq \mu_\text{max} (w^\circ - w^\bullet)^{\sf T}  [\nabla_{w^{\sf T}} p(w^\circ) - \nabla_{w^{\sf T}} p(w^\bullet)] \notag \\
& ~~~- t\mu_\text{max} 
\|w^\circ - w^\bullet\| \cdot \left\|\nabla_{w^{\sf T}} p(w^\circ) - \nabla_{w^{\sf T}} p(w^\bullet) \right\| \notag \\
& \geq - t\mu_\text{max} \delta_p \|w^\circ - w^\bullet\|^2 \end{align}    
where we used Assumptions~\ref{assump:strongmono} and~\ref{assump:Lipschitz}, the convexity property (\ref{conv_p}), and the Lipschitz continuous property (\ref{Lippenalfunc}).

\section{Proof of Theorem~\ref{CONVEG_GRADSTO}} \label{proof:conveg_gradsto}

We first note that assumption (\ref{gradnoise2}) can be rewritten as
\begin{align}
\label{gradnoise21}
\mathbb{E}\left[\|\bm{s}_i(\bm{w}_{i-1})\|^2| \bm{\mathcal{F}}_{i-1}\right] &\leq \alpha \|\bm{w}_{i-1} - w^\star + w^\star \|^2+\beta \notag \\
&\leq 2 \alpha \|\widetilde{\bm{w}}_{i-1}\|^2 + \beta'
\end{align}  
where we used $\|a+b\|^2 \leq 2\|a\|^2 + 2\|b\|^2$ and introduced $\beta' \triangleq \beta + 2\alpha \|w^\star\|^2$.	Then, using properties (\ref{gradnoise1}), (\ref{gradnoise2}), (\ref{combineLip}), (\ref{heterstrongmono}), and the fact that $F^p(w^{\star})=0$, we can express the mean-square error $\mathbb{E}\|\widetilde{\bm{w}}_i\|^2$ from (\ref{errStoGradient}) as
\begin{align}
\label{error_recursto}
&\mathbb{E}\|\widetilde{\bm{w}}_i\|^2 \notag \\ &=\mathbb{E}\|\widetilde{\bm{w}}_{i-1}\|^2 - 2 \;\mathbb{E}\left[\widetilde{\bm{w}}_{i-1}^{\sf T} U(F^p(w^\star)-F^p(\bm{w}_{i-1}))\right]  \notag \\
&~~~+ \mathbb{E} \|F^p(w^\star)-F^p(\bm{w}_{i-1})\|^2_{U^2} + \mathbb{E}\|\bm{s}_i(\bm{w}_{i-1})\|^2_{U^2}  \notag \\
& \leq \mathbb{E}\|\widetilde{\bm{w}}_{i-1}\|^2 - 2 \;\mathbb{E}\left[\widetilde{\bm{w}}_{i-1}^{\sf T} U(F^p(w^\star)-F^p(\bm{w}_{i-1}))\right] \notag \\
&~~~+ \mu_\text{max}^2 \mathbb{E} \|F^p(w^\star)-F^p(\bm{w}_{i-1})\|^2 + \mu_\text{max}^2 \mathbb{E}\|\bm{s}_i(\bm{w}_{i-1})\|^2
\notag \\
& \leq \left(1-2\mu_\text{max}\nu'+\mu_\text{max}^2 [(\delta + \rho\delta_p)^2+ 2\alpha] \right) \mathbb{E} \|\widetilde{\bm{w}}_{i-1}\|^2 + \mu_\text{max}^2  \beta 
\end{align}   
Note that from $\delta\geq \nu$ in (\ref{vdelta}) we have
\begin{align}
&1-2\mu_\text{max}\nu'+\mu_\text{max}^2 [(\delta + \rho\delta_p)^2+ 2\alpha] \notag \\
& = 1- 2 \mu_\text{max} [\nu - t (\delta+\rho \delta_p)]+\mu_\text{max}^2 [(\delta + \rho\delta_p)^2+ 2\alpha] \notag \\
&= (1-\mu_\text{max} \nu)^2 + \mu_\text{max}^2 ((\delta + \rho\delta_p)^2+ 2\alpha-\nu^2) \notag \\ 
&~~~ + 2 \mu_\text{max} t (\delta+\rho \delta_p) \geq 0
\end{align}  
Therefore, the mean-square error is stable asymptotically, as $i\rightarrow\infty$, when the step-size $\mu_\text{max}$ satisfies 
\begin{align}
& |1-2\mu_\text{max}\nu'+\mu_\text{max}^2 [(\delta + \rho\delta_p)^2+ 2\alpha]|<1 \notag \\
\Longleftrightarrow ~~~& -1 < 1-2\mu_\text{max}\nu'+\mu_\text{max}^2 [(\delta + \rho\delta_p)^2+ 2\alpha] < 1 \notag \\
\Longleftrightarrow ~~~& 0 < \mu_\text{max} < \frac{2\nu'}{(\delta + \rho\delta_p)^2+ 2\alpha} 
\end{align}  
when $\nu'$ is positive, i.e.,
\begin{align}
\label{condvprime}
\nu' = \nu-t(\delta+\rho \delta_p) >0 ~~\Longleftrightarrow~~ t < \frac{\nu}{\delta+\rho \delta_p}
\end{align}  
This leads to the conditions in (\ref{stepsize_Sto}), and the resulting mean-squared error is upper bounded by
\begin{align}
\lim\limits_{i \rightarrow \infty} \sup \mathbb{E}\|\widetilde{\bm{w}}_i\|^2 &\leq \frac{\mu_\text{max}  \beta}{2\nu' - \mu_\text{max} [(\delta + \rho\delta_p)^2+ 2\alpha]}= O(\mu_\text{max})
\end{align}  	

\section{Proof of Theorem~\ref{CONVEG_GRAD2}} \label{proof:conveg_grad2} 

Let us consider two unequal vectors $w_{i-1}^\circ$ and $w_{i-1}^\bullet$ with corresponding vectors $\{\phi_i^\circ, \psi_i^\circ, w_i^\circ\}$ and $\{\phi_i^\bullet, \psi_i^\bullet, w_i^\bullet\}$ in implementation (\ref{netGradient20})--(\ref{netGradient22}). 
The squared Euclidean distance between $\phi_i^\circ$ and $\phi_i^\bullet$ is given by
\begin{align}
\label{Gradient3first}
&\!\!\!\|\phi_i^\circ - \phi_i^\bullet\|^2 \notag \\
&\!\!\!= \|(w_{i-1}^\circ - w_{i-1}^\bullet) - c_1 \rho U [\nabla_{w^{\sf T}} p (w_{i-1}^\circ)-\nabla_{w^{\sf T}} p (w_{i-1}^\bullet)]\|^2 \notag \\
&\!\!\!\leq \|w_{i-1}^\circ - w_{i-1}^\bullet\|^2 + c_1^2 \mu_\text{max}^2 \rho^2 \|\nabla_{w^{\sf T}} p (w_{i-1}^\circ)-\nabla_{w^{\sf T}} p (w_{i-1}^\bullet)\|^2 \notag \\
&~~~- 2 c_1 \rho(w_{i-1}^\circ - w_{i-1}^\bullet)^{\sf T} U [\nabla_{w^{\sf T}} p (w_{i-1}^\circ)-\nabla_{w^{\sf T}} p (w_{i-1}^\bullet)] \notag \\
&\!\!\! \leq (1 + 2c_1 t \mu_\text{max} \rho \delta_p + c_1 \mu_\text{max}^2 \rho^2 \delta_p^2)\|w_{i-1}^\circ - w_{i-1}^\bullet\|^2
\end{align}   
where we used the properties (\ref{Lippenalfunc}), (\ref{heterpenalty}), and $c_1^2 =c_1$ from (\ref{ccc}). Using similar arguments we have
\begin{align}
\label{Gradient3second}
\|w_i^\circ - w_i^\bullet\|^2 \leq (1 + 2c_2 t \mu_\text{max} \rho \delta_p + c_2 \mu_\text{max}^2 \rho^2 \delta_p^2)\|\psi_i^\circ - \psi_i^\bullet\|^2
\end{align}   
For $\psi_i^\circ$ and $\psi_i^\bullet$, we can write
\begin{align}
\label{Gradient3third}
\|\psi_i^\circ - \psi_i^\bullet\|^2
& = \|(\phi_i^\circ - \phi_i^\bullet) - U (F(\phi_i^\circ)-F(\phi_i^\bullet))\|^2 \notag \\
&\leq \|\phi_i^\circ - \phi_i^\bullet\|^2 - 2 (\phi_i^\circ-\phi_i^\bullet)^{\sf T} U (F(\phi_i^\circ)-F(\phi_i^\bullet)) \notag \\
&~~~+ \mu_\text{max}^2 \|F(\phi_i^\circ)-F(\phi_i^\bullet)\|^2 \notag \\
& \leq (1-2\mu_\text{max} \nu'' + \mu_\text{max}^2 \delta^2)\|\phi_i^\circ-\phi_i^\bullet\|^2
\end{align}   
where we used (\ref{heterblock}) and Assumption~\ref{assump:Lipschitz}.
Combining (\ref{Gradient3first}), (\ref{Gradient3second}), and (\ref{Gradient3third}) we get
\begin{align}
\|w_i^\circ - w_i^\bullet\|^2 
&\leq (1 + 2c_1 t \mu_\text{max} \rho \delta_p + c_1 \mu_\text{max}^2 \rho^2 \delta_p^2) \notag \\
&~~~\times (1 + 2c_2 t \mu_\text{max} \rho \delta_p + c_2 \mu_\text{max}^2 \rho^2 \delta_p^2) \notag \\
&~~~\times (1-2\mu_\text{max} \nu'' + \mu_\text{max}^2 \delta^2)\|w_{i-1}^\circ - w_{i-1}^\bullet\|^2 \notag \\
&= (1 + 2 t \mu_\text{max} \rho \delta_p + \mu_\text{max}^2 \rho^2 \delta_p^2) \notag \\
&~~~\times(1-2\mu_\text{max} \nu'' + \mu_\text{max}^2 \delta^2) \notag \\
&~~~\times\|w_{i-1}^\circ - w_{i-1}^\bullet\|^2 
\end{align}   
The mapping $w_{i-1} \mapsto w_i$ is a contraction if
\begin{align}
\label{stepcond1}
&| (1 + 2 t \mu_\text{max} \rho \delta_p + \mu_\text{max}^2 \rho^2 \delta_p^2) (1-2\mu_\text{max} \nu'' + \mu_\text{max}^2 \delta^2)| <1 \notag \\
&\Longleftrightarrow  -1 <  (1 + 2 t \mu_\text{max} \rho \delta_p + \mu_\text{max}^2 \rho^2 \delta_p^2) \notag \\
&\qquad \qquad~~~\times (1-2\mu_\text{max} \nu'' + \mu_\text{max}^2 \delta^2) <1
\end{align}  
We note that 
\begin{align}
\label{alwpos}
1-&2\mu_\text{max} \nu'' + \mu_\text{max}^2 \delta^2 \notag \\
&= (1-\mu_\text{max} \nu)^2 + \mu_\text{max}^2 (\delta^2 -\nu^2) + 2 \mu_\text{max} t \delta \geq 0
\end{align}  
Therefore, the inequality on the left-hand side of (\ref{stepcond1}) always holds. Expanding the product of the two terms and using $\nu'=\nu''-t\rho\delta_p$ we get for the inequality on the right-hand side of (\ref{stepcond1}) that we must have 
\begin{align}
\label{stepcond2}
1-a_1 <1 \Longleftrightarrow a_1 > 0
\end{align}  
where
\begin{align}
\label{a_1}
a_1 &\triangleq 2\mu_\text{max} \nu' - \mu_\text{max}^2 (\delta^2+\rho^2 \delta_p^2-4t\nu''\rho\delta_p) \notag \\
&~~~ +\mu_\text{max}^3 (\rho^2 \delta_p^2 \nu''-t \rho \delta_p \delta^2) - \mu_\text{max}^4 \rho^2 \delta_p^2 \delta^2 
\end{align}  
Therefore, if we can guarantee
\begin{align}
\label{cond1}
\nu' &>0 \\
\label{cond2}
\delta^2+\rho^2 \delta_p^2-4t\nu''\rho\delta_p&>0  \\
\label{cond3}
\mu_\text{max}^3 (\rho^2 \delta_p^2 \nu''-t \rho \delta_p \delta^2) &> \mu_\text{max}^4 \rho^2 \delta_p^2 \delta^2 
\end{align}  
then the condition $a_1>0$ is satisfied if 
\begin{align}
2\mu_\text{max} \nu' - \mu_\text{max}^2 (\delta^2+\rho^2 \delta_p^2-4t\nu''\rho\delta_p)>0
\end{align}  
which means
\begin{align}
\label{step1}
\mu_\text{max} < 2\nu'/(\delta^2+\rho^2 \delta_p^2-4t\nu''\rho\delta_p)
\end{align}  
Let us examine conditions (\ref{cond1})--(\ref{cond3}). From  (\ref{condvprime}) we know that the condition of a positive $\nu'$ holds if
\begin{align}
\label{condit}
t < \nu/(\delta+\rho \delta_p)
\end{align}  
For the second condition (\ref{cond2}), we now show that if $\rho$ is sufficiently large such that
\begin{align}
\label{condirho}
\rho > \delta / \delta_p ~~\Longleftrightarrow~~ \rho \delta_p > \delta
\end{align}  
then
\begin{align}
f(t) &\triangleq \delta^2+\rho^2 \delta_p^2-4t\nu''\rho\delta_p \notag \\
&=4t^2 \delta \rho\delta_p - 4t \nu \rho \delta_p + \delta^2 + \rho^2 \delta_p^2 >\delta^2 >0 
\end{align}  
where we used $\nu''=\nu- t\delta$. Note that $f(t)$ is a quadratic function of $t$ and has a minimum at $t^o=\nu/(2\delta)$. Therefore, it is required that
\begin{align}
\label{ftmin}
&f(t^o) = \delta^2 + \rho^2 \delta_p^2 - \rho \delta_p \cdot \nu^2/\delta  >\delta^2 
~\Longleftrightarrow~  \rho \delta_p > \nu^2/\delta
\end{align}  
Under condition (\ref{condirho}) and from the fact $\delta\geq \nu$, we get
\begin{align}
\delta \rho \delta_p > \delta^2 \geq \nu^2 
&\Longrightarrow  
\rho \delta_p > \nu^2/\delta
\end{align}  
which ensures $f(t)>\delta^2>0$ for any $t$. For the third condition (\ref{cond3}), we first note that we need the left-hand side of (\ref{cond3}) to be positive, which requires
\begin{align}
\label{midcond}
\rho^2 \delta_p^2 \nu''-t \rho \delta_p \delta^2 >0 ~~\Longleftrightarrow~~ \rho \delta_p \nu' + t(\rho^2 \delta_p^2-\delta^2)>0
\end{align}  
where we used $\nu'' = \nu' + t\rho\delta_p$. By (\ref{condirho}) and (\ref{cond1}) we know that (\ref{midcond}) holds. Then, we have
\begin{align}
\label{condstep2}
&\rho^2 \delta_p^2 \nu''-t \rho \delta_p \delta^2 > \mu_\text{max}\rho^2 \delta_p^2 \delta^2 \notag \\
\Longleftrightarrow ~~& \rho \delta_p \nu' + t(\rho^2 \delta_p^2-\delta^2) > \mu_\text{max} \rho \delta_p \delta^2 \notag \\
\Longleftrightarrow ~~& \mu_\text{max} < \left(\nu' +  \frac{t(\rho^2\delta_p^2-\delta^2)}{\rho\delta_p}\right)/{\delta^2}
\end{align}  
Therefore, we arrive at the following sufficient conditions for the convergence of (\ref{netGradient20})--(\ref{netGradient22}):
\begin{align}
0<\mu_\text{max} < \mu_o , \quad t < \frac{\nu}{\delta+\rho \delta_p}, \quad \rho > \frac{\delta}{\delta_p}
\end{align}   
where 
\begin{align}
\mu_o \triangleq \min\left\{\frac{2\nu'}{\delta^2+\rho^2 \delta_p^2-4t\nu''\rho\delta_p} ,\frac{\nu' +  \frac{t(\rho^2\delta_p^2-\delta^2)}{\rho\delta_p}}{\delta^2}\right\}
\end{align}  

\section{Proof of Theorem~\ref{CONVEG_GRADSTO2}} \label{proof:conveg_gradsto2}

Subtracting (\ref{netStoGradient20})--(\ref{netStoGradient22}) from (\ref{fixGradient0})--(\ref{fixGradient2}) we get
\begin{align}
\label{neterr0}
\widetilde{\bm{\phi}}_i^\infty &= \widetilde{\bm{w}}_{i-1}^\infty - c_1 \rho U \left[\nabla_{w^{\sf T}} p(w^\infty)-\nabla_{w^{\sf T}} p(\bm{w}_{i-1})\right]\\
\label{neterr1}
\widetilde{\bm{\psi}}_i^\infty &= \widetilde{\bm{\phi}}_i^\infty - U \left[F(\phi^\infty)-F(\bm{\phi}_i)\right] + U \bm{s}_i (\bm{\phi}_i) \\
\label{neterr2}
\widetilde{\bm{w}}_i^\infty &= \widetilde{\bm{\psi}}_i^\infty - c_2 \rho U \left[\nabla_{w^{\sf T}} p(\psi^\infty)-\nabla_{w^{\sf T}} p(\bm{\psi}_i)\right]
\end{align}  
where $\widetilde{\bm{\phi}}^\infty_i \triangleq \phi^\infty - \bm{\phi}_i$, $\widetilde{\bm{\psi}}^\infty_i \triangleq \psi^\infty - \bm{\psi}_i$, and $\widetilde{\bm{w}}^\infty_i \triangleq w^\infty - \bm{w}_i$. From (\ref{neterr0}) we have
\begin{align}
\label{neterr0t}
&\mathbb{E}\|\widetilde{\bm{\phi}}_i^\infty\|^2 \notag \\
& \leq \mathbb{E}\|\widetilde{\bm{w}}_{i-1}^\infty\|^2 + c_1^2 \mu_\text{max}^2 \rho^2 \mathbb{E}\|\nabla_{w^{\sf T}} p(w^\infty) - \nabla_{w^{\sf T}} p(\bm{w}_{i-1})\|^2 \notag \\
& ~~~ - 2 c_1 \rho \mathbb{E} \left[\widetilde{\bm{w}}_{i-1}^{\infty{\sf T}} U \left[\nabla_{w^{\sf T}} p(w^\infty) - \nabla_{w^{\sf T}} p(\bm{w}_{i-1})\right]\right] \notag \\
& \leq (1+2c_1 t\mu_\text{max} \rho \delta_p + c_1 \mu_\text{max}^2 \rho^2 \delta_p^2) \mathbb{E}\|\widetilde{\bm{w}}_{i-1}^\infty\|^2
\end{align}  
and, similarly, from (\ref{neterr2}) we obtain
\begin{align}
\label{neterr2t}
\mathbb{E}\|\widetilde{\bm{w}}_i^\infty\|^2 & \leq (1+ 2c_2 t\mu_\text{max} \rho \delta_p+ c_2 \mu_\text{max}^2 \rho^2 \delta_p^2) \mathbb{E}\|\widetilde{\bm{\psi}}_i^\infty\|^2
\end{align}  
Similar to (\ref{gradnoise21}), we can rewrite assumption (\ref{gradnoise2}) as
\begin{align}
\label{gradnoise22}
\mathbb{E}\left[\|\bm{s}_i(\bm{w}_{i-1})\|^2| \bm{\mathcal{F}}_{i-1}\right] 
&\leq 2 \alpha \|\widetilde{\bm{w}}^\infty_{i-1}\|^2 + \beta''
\end{align}  
for $\beta'' \triangleq \beta + 2\alpha \|w^\infty\|^2$.
Then, from (\ref{neterr1}) we obtain:
\begin{align}
\label{neterr1t}
\mathbb{E} \|\widetilde{\bm{\psi}}_i^\infty\|^2 
&\leq \mathbb{E}\|\widetilde{\bm{\phi}}_i^\infty\|^2 + \mu_\text{max}^2 \mathbb{E}\|F(\phi^\infty)-F(\bm{\phi}_i) \|^2 \notag \\
&~~~+ \mu_\text{max}^2 \mathbb{E} \|\bm{s} (\bm{\phi}_i)\|^2 - 2 \mathbb{E}\left[\widetilde{\bm{\phi}}_i^{\infty{\sf T}}U [F(\phi^\infty)-F(\bm{\phi}_i)] \right] \notag \\
&\leq \left(1 - 2\mu_\text{max} \nu'' + \mu_\text{max}^2 (\delta^2 + 2\alpha) \right) \mathbb{E}\|\widetilde{\bm{\phi}}_i^\infty\|^2 + \mu_\text{max}^2 \beta''
\end{align}  
Therefore, we can combine (\ref{neterr0t})--(\ref{neterr1t}) to get
\begin{align}
\label{MSErecur}
&\mathbb{E}\|\widetilde{\bm{w}}_i^\infty\|^2 \notag \\
&\leq (1 + 2c_1 t\mu_\text{max} \rho \delta_p + c_1 \mu_\text{max}^2 \rho^2 \delta_p^2) \notag \\
&~~~\times (1 + 2c_2 t\mu_\text{max} \rho \delta_p + c_2 \mu_\text{max}^2 \rho^2 \delta_p^2)  \notag \\
&~~~\times \left(1 - 2\mu_\text{max} \nu'' + \mu_\text{max}^2 (\delta^2 + 2\alpha) \right) \mathbb{E}\|\widetilde{\bm{w}}_{i-1}^\infty\|^2 \notag \\
&~~~+ \mu_\text{max}^2 (1+ 2c_2 t\mu_\text{max} \rho \delta_p + c_2 \mu_\text{max}^2 \rho^2 \delta_p^2) \beta'' \notag \\
&=(1 + 2 t\mu_\text{max} \rho \delta_p + \mu_\text{max}^2 \rho^2 \delta_p^2) \left(1 - 2\mu_\text{max} \nu'' + \mu_\text{max}^2 (\delta^2 + 2\alpha) \right) \notag \\
&~~~ \times \mathbb{E}\|\widetilde{\bm{w}}_{i-1}^\infty\|^2 + \mu_\text{max}^2 (1+ 2c_2 t\mu_\text{max} \rho \delta_p + c_2 \mu_\text{max}^2 \rho^2 \delta_p^2) \beta''
\end{align}  
We expand the product of the two terms as
\begin{align}
(1 &+ 2 t\mu_\text{max} \rho \delta_p + \mu_\text{max}^2 \rho^2 \delta_p^2) (1 - 2\mu_\text{max} \nu'' + \mu_\text{max}^2 (\delta^2 + 2\alpha) ) \notag \\
& \triangleq 1-a_2
\end{align}  
where 
\begin{align}
a_2 &\triangleq 2\mu_\text{max} \nu' - \mu_\text{max}^2 (\delta^2+2\alpha+\rho^2 \delta_p^2-4t\nu''\rho\delta_p) \notag \\
&~~~ +\mu_\text{max}^3 (\rho^2 \delta_p^2 \nu''-t \rho \delta_p (\delta^2+2\alpha)) - \mu_\text{max}^4 \rho^2 \delta_p^2 (\delta^2+2\alpha) 
\end{align}  
Then, the mean-square error $\mathbb{E}\|\widetilde{\bm{w}}_i^\infty\|^2$ converges asymptotically as $i \rightarrow \infty$ if we have $|1-a_2|<1$, which requires $a_2>0$ since from (\ref{alwpos}) we know $1-a_2\geq 0$. 
Following a similar argument to the one presented in Appendix~\ref{proof:conveg_grad2}, we obtain that the following conditions ensure the convergence of $\mathbb{E}\|\widetilde{\bm{w}}_i^\infty\|^2$:
\begin{align}
&\nu' >0 \\
&\delta^2 + 2\alpha +\rho^2 \delta_p^2-4t\nu''\rho\delta_p>0  \\
&\mu_\text{max}^3 (\rho^2 \delta_p^2 \nu''-t \rho \delta_p (\delta^2+2\alpha)) > \mu_\text{max}^4 \rho^2 \delta_p^2 (\delta^2+2\alpha) \\
\label{condl2}
&2\mu_\text{max} \nu' - \mu_\text{max}^2 (\delta^2+2\alpha+\rho^2 \delta_p^2-4t\nu''\rho\delta_p)>0
\end{align}  
The first two yield the same results in (\ref{condit}) and (\ref{condirho}), i.e., 
\begin{align}
\label{newtrho}
t < \frac{\nu}{\delta+\rho \delta_p}, \qquad 
\rho > \frac{\delta}{\delta_p} 
\end{align}   
For the third condition we need to ensure 
\begin{align}
&\rho^2 \delta_p^2 \nu''-t \rho \delta_p (\delta^2+2\alpha) >0 \notag \\
\Longleftrightarrow~~ &\rho \delta_p \nu' + t(\rho^2 \delta_p^2-(\delta^2+2\alpha))>0
\end{align}  
A stricter condition on $\rho$ is therefore required:
\begin{align}
\label{newcondirho}
\rho > \frac{\sqrt{\delta^2+2\alpha}}{\delta_p} 
\end{align}   
We then get
\begin{align}
&\rho \delta_p \nu''-t (\delta^2+2\alpha) > \mu_\text{max} \rho \delta_p (\delta^2+2\alpha) \notag \\
\Longleftrightarrow~~ & \mu_\text{max} < \frac{\nu' +  \frac{t(\rho^2\delta_p^2-(\delta^2+2\alpha))}{\rho\delta_p}}{\delta^2+2\alpha}
\end{align}  
Combining the last condition (\ref{condl2}), we get the step-size condition as
\begin{align}
\label{newcondu}
0<\mu_\text{max} < \mu'_o
\end{align}   
where 
\begin{align}
\mu'_o \triangleq \min\!\Bigg\{\!\frac{2\nu'}{\delta^2+2\alpha+\rho^2 \delta_p^2-4t\nu''\rho\delta_p}, \frac{\nu' +  \frac{t(\rho^2\delta_p^2-(\delta^2+2\alpha))}{\rho\delta_p}}{\delta^2+2\alpha} \!\Bigg\}
\end{align}  
Therefore, under conditions (\ref{newtrho}), (\ref{newcondirho}), and (\ref{newcondu}), the recursion (\ref{MSErecur}) is stable and the resulting mean-square error is upper bounded by
\begin{align}
\lim\limits_{i \rightarrow \infty} \sup \mathbb{E}\|\widetilde{\bm{w}}_i^\infty\|^2 &\leq \frac{\mu_\text{max}^2 (1+ 2c_2 t\mu_\text{max} \rho \delta_p + c_2 \mu_\text{max}^2 \rho^2 \delta_p^2) \beta''}{a_2} \notag \\
&=\frac{\mu_\text{max} (1+ 2c_2 t\mu_\text{max} \rho \delta_p + c_2 \mu_\text{max}^2 \rho^2 \delta_p^2) \beta''}{2\nu' - O(\mu_\text{max})} \notag \\
& = O(\mu_\text{max})
\end{align}  
for sufficiently small step-sizes.	

\section{Proof of Theorem~\ref{SMALLBIAS}} \label{proof:smallbias}

We recall from (\ref{NashCond}) that the Nash equilibrium $w^{\star}$ satisfies the relation:
\begin{align}
\label{pseudoRecur0}
w^\star &= w^\star - U \left[F(w^\star) + \rho \nabla_{w^{\sf T}} p(w^\star)\right] \notag \\
&= w^\star - U F(w^\star) - c_1 \rho U \nabla_{w^{\sf T}} p(w^\star) - c_2 \rho U \nabla_{w^{\sf T}} p(w^\star) \notag \\
&= \phi^\star - U F(w^\star) - c_2 \rho U \nabla_{w^{\sf T}} p(w^\star) \notag \\
&= \psi^\star - c_2 \rho U \nabla_{w^{\sf T}} p(w^\star)
\end{align}  
where we introduced two auxiliary variables $\phi^\star$ and $\psi^\star$:
\begin{align}
\label{pseudoRecur1}
\phi^\star &= w^\star - c_1 \rho U \nabla_{w^{\sf T}} p(w^\star) \\
\label{pseudoRecur2}
\psi^\star &= \phi^\star - U F(w^\star)
\end{align}  
If we further introduce the error vectors $\widetilde{\phi} \triangleq \phi^\star-\phi^\infty$, $\widetilde{\psi} \triangleq \psi^\star-\psi^\infty$, and $\widetilde{w} \triangleq w^\star-w^\infty$, then using (\ref{pseudoRecur0})--(\ref{pseudoRecur2}) we have 
\begin{align}
\label{bias30}
\widetilde{\phi} &= \widetilde{w} - c_1 \rho U [\nabla_{w^{\sf T}} p(w^\star)-\nabla_{w^{\sf T}} p(w^\infty)] \\
\label{bias31}
\widetilde{\psi} &= \widetilde{\phi} - U \left[F(w^\star)-F(\phi^\infty)\right]\\
\label{bias32}
\widetilde{w} &= \widetilde{\psi} - c_2 \rho U [\nabla_{w^{\sf T}} p(w^\star)-\nabla_{w^{\sf T}} p(\psi^\infty)]
\end{align}  
From (\ref{bias30}), the squared norm of $\widetilde{\phi}$ satisfies
\begin{align}
\label{result0}
\|\widetilde{\phi}\|^2 &\leq \|\widetilde{w}\|^2 - 2 c_1 \rho \widetilde{w}^{\sf T} U [\nabla_{w^{\sf T}} p(w^\star)-\nabla_{w^{\sf T}} p(w^\infty)] \notag \\
&~~~+ c_1 \mu_\text{max}^2 \rho^2  \|\nabla_{w^{\sf T}} p(w^\star)-\nabla_{w^{\sf T}} p(w^\infty)\|^2 \notag \\
& \leq \mathcal{Y}_1 \|\widetilde{w}\|^2 
\end{align}  
where we used (\ref{Lippenalfunc}) and (\ref{heterpenalty}) and introduced 
\begin{align}
\mathcal{Y}_1 \triangleq 
1+2 c_1 t \mu_\text{max} \rho \delta_p + c_1 \mu_\text{max}^2 \rho^2 \delta_p^2
\end{align}  
From (\ref{bias31}), the squared norm of $\widetilde{\psi}$ satisfies
\begin{align}
\|\widetilde{\psi}\|^2 &= \|\widetilde{\phi}\|^2 - 2 \widetilde{\phi}^{\sf T} U [F(w^\star)-F(\phi^\infty)] \notag \\
&~~~ + \|U[F(w^\star)-F(\phi^\infty)]\|^2
\end{align}  
We note that 
\begin{align}
\label{psistep1}
- 2& \widetilde{\phi}^{\sf T} U [F(w^\star)-F(\phi^\infty)] \notag \\
&= - 2 \widetilde{\phi}^{\sf T} U [F(\phi^\star)-F(\phi^\infty)] - 2 \widetilde{\phi}^{\sf T} U [F(w^\star)-F(\phi^\star)] \notag \\
&\stackrel{(a)}{\leq} -2 \mu_\text{max} \nu'' \|\widetilde{\phi}\|^2 + 2 \mu_\text{max} \|F(w^\star)-F(\phi^\star)\| \cdot \|\widetilde{\phi}\| \notag \\
&\stackrel{(b)}{\leq} -2 \mu_\text{max} \nu'' \|\widetilde{\phi}\|^2 + 2 c_1 \mu^2_\text{max}  \delta \|F(w^\star)\| \cdot \|\widetilde{\phi}\| 
\end{align}  
where step (a) is from (\ref{heterblock}) and H{\"o}lder's inequality and step (b) is due to
\begin{align}
\|F(w^\star)-F(\phi^\star)\| &\leq \delta \|w^\star-\phi^\star\| \leq c_1 \mu_\text{max} \delta \|F(w^\star)\| 
\end{align}  
since from (\ref{pseudoRecur1}) and (\ref{NashCond}) we have 
\begin{align}
\|w^\star-\phi^\star\|
&=\|c_1 \rho U \nabla_{w^{\sf T}} p(w^\star)\| \leq  c_1 \mu_\text{max} \|F(w^\star)\| 
\end{align}  
We further note that 
\begin{align}
\label{psistep2}
\|U[F(w^\star)-F(\phi^\infty)]\|^2 &\leq \mu^2_\text{max} \delta^2 \|w^\star - \phi^\infty \|^2 \notag \\
& \leq \mu^2_\text{max} \delta^2 \|\widetilde{\phi} \|^2 + 2 c_1  \mu^3_\text{max} \delta^2 \|F(w^\star)\| \cdot \|\widetilde{\phi} \| \notag \\
&~~~+ c_1 \mu^4_\text{max} \delta^2\| F(w^\star)\|^2
\end{align}  
where we used the fact $w^\star - \phi^\infty = \phi^\star - \phi^\infty + w^\star - \phi^\star$ and 
\begin{align}
&\|w^\star-\phi^\infty\|^2 \notag \\
&= \|\widetilde{\phi} \|^2 + 2\widetilde{\phi}^{\sf T} (w^\star - \phi^\star) + \|w^\star - \phi^\star\|^2 \notag \\
&\leq \|\widetilde{\phi} \|^2 + 2 \|w^\star - \phi^\star\| \cdot \|\widetilde{\phi} \| + 
c_1 \mu^2_\text{max}\| F(w^\star)\|^2 \notag \\
& \leq \|\widetilde{\phi} \|^2 + 2 c_1 \mu_\text{max} \|F(w^\star)\| \cdot \|\widetilde{\phi} \|+ 
c_1 \mu^2_\text{max}\| F(w^\star)\|^2
\end{align}  
Using (\ref{psistep1}) and (\ref{psistep2}) we get
\begin{align}
\label{result1}
\|\widetilde{\psi}\|^2 &\leq \mathcal{X}  \|\widetilde{\phi}\|^2 + 2 c_1 \mu^2_\text{max} (1+\mu_\text{max}\delta)  \delta \|F(w^\star)\| \cdot \|\widetilde{\phi}\| \notag \\
&~~~+ c_1 \mu^4_\text{max} \delta^2\| F(w^\star)\|^2 
\end{align}  
where we introduced 
\begin{align}
\mathcal{X} \triangleq 1-2\mu_\text{max} \nu''+ \mu^2_\text{max} \delta^2 \geq 0
\end{align}  
Note that $\mathcal{X}$ is always nonnegative by (\ref{alwpos}).
Similarly, from (\ref{bias32}) we have
\begin{align}
\label{psi0}
\|\widetilde{w}\|^2 &\leq \|\widetilde{\psi}\|^2 + c_2 \mu_\text{max}^2 \rho^2  \|\nabla_{w^{\sf T}} p(w^\star)-\nabla_{w^{\sf T}} p(\psi^\infty)\|^2 \notag \\
&~~~- 2 c_2 \rho \widetilde{\psi}^{\sf T} U [\nabla_{w^{\sf T}} p(w^\star)-\nabla_{w^{\sf T}} p(\psi^\infty)] \notag \\
& \leq \|\widetilde{\psi}\|^2 + c_2 \mu_\text{max}^2 \rho^2 \delta_p^2 \|w^\star-\psi^\infty\|^2 \notag \\
&~~~- 2 c_2 \rho \widetilde{\psi}^{\sf T} U [\nabla_{w^{\sf T}} p(w^\star)-\nabla_{w^{\sf T}} p(\psi^\star)] \notag \\
&~~~+ 2c_2 t \mu_\text{max} \rho \delta_p \|\widetilde{\psi}\|^2
\end{align}  
where we rewrote 
\begin{align}
&\nabla_{w^{\sf T}} p(w^\star)-\nabla_{w^{\sf T}} p(w^\infty) \notag \\
&= \nabla_{w^{\sf T}} p(w^\star)-\nabla_{w^{\sf T}} p(\psi^\star)+\nabla_{w^{\sf T}} p(\psi^\star)-\nabla_{w^{\sf T}} p(w^\infty)
\end{align}  
and used (\ref{Lippenalfunc}) and (\ref{heterpenalty}).
By (\ref{pseudoRecur0}) we know that 
\begin{align}
\label{bias1}
\|w^\star-\psi^\star\| 
&=\|c_2 \rho U \nabla_{w^{\sf T}} p(w^\star)\| \notag \\
&\leq c_2 \mu_\text{max} \|\rho \nabla_{w^{\sf T}} p(w^\star)\| \notag \\
&= c_2 \mu_\text{max} \|F(w^\star)\| 
\end{align}  
Then, it follows that
\begin{align}
\label{psi1}
&\|w^\star-\psi^\infty\|^2 \notag \\
&= \|\widetilde{\psi} \|^2 + 2\widetilde{\psi}^{\sf T} (w^\star - \psi^\star) + \|w^\star - \psi^\star\|^2 \notag \\
&\leq \|\widetilde{\psi} \|^2 + 2 \|w^\star - \psi^\star\| \cdot \|\widetilde{\psi} \| + 
c_2 \mu^2_\text{max}\| F(w^\star)\|^2 \notag \\
& \leq \|\widetilde{\psi} \|^2 + 2 c_2 \mu_\text{max} \|F(w^\star)\| \cdot \|\widetilde{\psi} \|+ 
c_2 \mu^2_\text{max}\| F(w^\star)\|^2 
\end{align}  
Furthermore, we can use the Cauchy-Schwartz inequality and the Lipschitz-continuous assumption again to write
\begin{align}
\label{psi2}
- 2 c_2 &\rho \widetilde{\psi}^{\sf T} U [\nabla_{w^{\sf T}} p(w^\star)-\nabla_{w^{\sf T}} p(\psi^\star)] \notag \\
&\leq 2 c_2 \rho \mu_\text{max}  \|\nabla_{w^{\sf T}} p(w^\star)-\nabla_{w^{\sf T}} p(\psi^\star)\| \cdot \|\widetilde{\psi}\| \notag \\
&\leq 2 c_2 \mu_\text{max} \rho \delta_p   \|w^\star-\psi^\star\| \cdot \|\widetilde{\psi}\|\notag \\
&\leq 2 c_2 \mu_\text{max}^2 \rho \delta_p \|F(w^\star)\| \cdot  \|\widetilde{\psi}\| 
\end{align}  
where the last inequality is by (\ref{bias1}).
Substituting (\ref{psi1}) and (\ref{psi2}) into (\ref{psi0}), we get
\begin{align}
\label{result2}
\|\widetilde{w}\|^2 
& \leq \mathcal{Y}_2 \|\widetilde{\psi}\|^2 \notag + 2 c_2 \mu_\text{max}^2 (1+\mu_\text{max} \rho \delta_p) \rho \delta_p \|F(w^\star)\| \cdot \| \widetilde{\psi}\| \\
&~~~+ c_2\mu_\text{max}^4 \rho^2 \delta_p^2 \| F (w^\star)\|^2
\end{align}  
where we introduced 
\begin{align}
\mathcal{Y}_2 \triangleq 
1+2 c_2 t \mu_\text{max} \rho \delta_p + c_2 \mu_\text{max}^2 \rho^2 \delta_p^2
\end{align}  
To continue, we note the following properties:
\begin{align}
\mathcal{Y}_1  \mathcal{Y}_2 &= 1+2 t \mu_\text{max} \rho \delta_p + \mu_\text{max}^2 \rho^2 \delta_p^2 \triangleq \mathcal{Y} \\
c_1 \mathcal{Y}_2 &= c_1, \qquad~~ c_2 \mathcal{Y}_1 = c_2 \mathcal{Y} \\ 
c_1 \mathcal{Y}_1 &= c_1 \mathcal{Y}, \qquad c_2 \mathcal{Y}_2 = c_2 \mathcal{Y} \\
c_2 \|\widetilde{\psi}\|^2 &= c_2 \mathcal{X} \|\widetilde{w}\|^2 ~~\Longleftrightarrow~~ c_2 \|\widetilde{\psi}\| = c_2 \sqrt{\mathcal{X}} \|\widetilde{w}\|
\end{align}  
by recalling $c_1 \cdot c_2=1$ and $c_1+c_2=1$ in (\ref{ccc}).
Combining (\ref{result0}), (\ref{result1}) and (\ref{result2}) we obtain
\begin{align}
\label{allresult}
\|\widetilde{w}\|^2  
& \leq \mathcal{Y}_1 \mathcal{Y}_2 \mathcal{X} \|\widetilde{w}\|^2 \notag \\
&~~~+ 2 c_1 \mu_\text{max}^2 (1+\mu_\text{max} \delta) \delta \|F(w^\star)\| \sqrt{\mathcal{Y}_1} \cdot \| \widetilde{w}\| \notag \\
&~~~+ 2 c_2 \mu_\text{max}^2 (1+\mu_\text{max} \rho \delta_p) \rho \delta_p \|F(w^\star)\| \sqrt{\mathcal{Y}_2} \cdot \| \widetilde{w}\| \notag \\
&~~~+ c_1\mu_\text{max}^4 \delta^2 \| F (w^\star)\|^2 + c_2\mu_\text{max}^4 \rho^2 \delta_p^2 \| F (w^\star)\|^2 \notag \\
&= \mathcal{Y} \mathcal{X} \|\widetilde{w}\|^2 + 2 \mu_\text{max}^2 \|F(w^\star)\| \sqrt{\mathcal{Y}} \mathcal{Z} \cdot \| \widetilde{w}\|  \notag \\
&~~~+ \mu_\text{max}^4 (c_1\delta^2 + c_2 \rho^2 \delta_p^2) \cdot\| F (w^\star)\|^2
\end{align}  
where 
\begin{align}
\mathcal{Z} \triangleq c_1 (1+\mu_\text{max} \delta) \delta + c_2 (1+\mu_\text{max} \rho \delta_p) \rho \delta_p
\end{align}  
Noting $1-\mathcal{Y} \mathcal{X}=a_1$ as defined in (\ref{a_1}), we can rewrite (\ref{allresult}) as
\begin{align}
\label{squareab}
a_1 \|\widetilde{w}\|^2 - 2 b \| \widetilde{w}\| \leq \eta
\end{align}  
where 
\begin{align}
\label{defb}
b &\triangleq 2 \mu_\text{max}^2  \|F(w^\star)\| \sqrt{\mathcal{Y}} \mathcal{Z} \geq 0 \\
\label{defeta}
\eta &\triangleq \mu_\text{max}^4 (c_1\delta^2 + c_2 \rho^2 \delta_p^2) \cdot\| F (w^\star)\|^2 \geq 0 
\end{align}   
From Appendix~\ref{proof:conveg_grad2}, we know that $a_1 > 0$ if
\begin{align}
0<\mu_\text{max} < \mu_o, \quad t < \frac{\nu}{\delta+\rho \delta_p}, \quad \rho > \frac{\delta}{\delta_p}
\end{align}   
Under these conditions we can rewrite (\ref{squareab}) as
\begin{align}
& \left(\|\widetilde{w}\| - \frac{b}{a_1}\right)^2 \leq \frac{ \eta}{a_1} + \frac{b^2}{a_1^2} \notag \\
\Longleftrightarrow~~& \frac{b}{a_1} - \sqrt{\frac{\eta}{a_1}+\frac{b^2}{a_1^2}} \leq  \|\widetilde{w}\| \leq \frac{b}{a_1} + \sqrt{\frac{\eta}{a_1}+\frac{b^2}{a_1^2}}
\end{align}  
Noting that
\begin{align}
\frac{b}{a_1} - \sqrt{\frac{\eta}{a_1}+\frac{b^2}{a_1^2}} = \frac{b}{a_1} - \frac{\sqrt{b^2+ a_1 \eta}}{a_1} \leq 0
\end{align}  
we get 
\begin{align}
0 \leq  \|\widetilde{w}\| \leq \frac{b}{a_1} + \sqrt{\frac{\eta}{a_1}+\frac{b^2}{a_1^2}}
\end{align}  
Our goal is to study the bias performance for sufficiently small step-sizes, which can be examined from
\begin{align}
\lim\limits_{\mu_\text{max} \rightarrow 0} \sup \frac{\|\widetilde{w}\|}{\mu_\text{max}} \leq \lim\limits_{\mu_\text{max} \rightarrow 0} \frac{b}{a_1\mu_\text{max}} + \lim\limits_{\mu_\text{max} \rightarrow 0} \sqrt{\frac{\eta}{a_1\mu_\text{max}^2}+\frac{b^2}{a_1^2 \mu_\text{max}^2}}
\end{align}  
From (\ref{defb}) and (\ref{a_1}) we have  
\begin{align}
&\lim\limits_{\mu_\text{max} \rightarrow 0} \frac{b}{a_1\mu_\text{max}} \notag \\
&= \lim\limits_{\mu_\text{max} \rightarrow 0} \frac{2 \|F(w^\star)\| \sqrt{\mathcal{Y}} \mathcal{Z}}{2 \nu' - \mu_\text{max} (\delta^2 +  \rho^2 \delta_p^2-4t \nu'' \rho \delta_p) + O(\mu_\text{max}^2)} \notag \\
&= d_1 (c_1 \delta + c_2 \rho \delta_p)  
\end{align}  
where we used the fact $\lim\limits_{\mu_\text{max} \rightarrow 0} \mathcal{Y} = 1$ and introduced
\begin{align}
d_1 \triangleq \|F(w^\star)\|/\nu'   
\end{align}  
From the definition (\ref{defeta}) we get
\begin{align}
&\lim\limits_{\mu_\text{max} \rightarrow 0} \frac{\eta}{a_1\mu_\text{max}^2} \notag \\
&= \lim\limits_{\mu_\text{max} \rightarrow 0} \frac{  \mu_\text{max} (c_1\delta^2 + c_2 \rho^2 \delta_p^2) \cdot\| F (w^\star)\|^2 }{2 \nu' - \mu_\text{max} (\delta^2 +  \rho^2 \delta_p^2-4t \nu'' \rho \delta_p)+O(\mu_\text{max}^2)} \notag \\ 
&= 0  
\end{align}  
Consequently, we have
\begin{align}
\lim\limits_{\mu_\text{max} \rightarrow 0} \sup \frac{\|\widetilde{w}\|}{\mu_\text{max}} \leq 2d_1 (c_1 \delta + c_2 \rho \delta_p) < 2d_1 \rho \delta_p
\end{align}  
where we used the condition $\rho > \delta/\delta_p$ and the fact $c_1+c_2=1$.

\bibliographystyle{IEEEtran}
\bibliography{IEEEabrv,refs}

\end{document}